\documentclass[12pt,preprint]{aastex}
\usepackage{natbib}
\usepackage{longtable}
\usepackage[small]{caption}
\usepackage{lscape}
\usepackage{mathrsfs}
\usepackage{url}

\newcommand{\kms}{\ifmmode {\rm km~s}^{-1} \else km~s$^{-1}$\fi}
\newcommand{\ergs}{\ifmmode {\rm erg~ s}^{-1} \else erg~s$^{-1}$\fi}
\newcommand{\ergscm}{\ifmmode {\rm erg~s}^{-1} \else erg~s$^{-1}$ cm$^{-2}$\fi}
\newcommand{\Msun}{\ifmmode {\rm M}_{\odot} \else M$_{\odot}$\fi }
\newcommand{\Lsun}{\ifmmode {\rm L}_{\odot} \else L$_{\odot}$\fi}
\newcommand{\qo}{\ifmmode q_{\rm o} \else $q_{\rm o}$\fi}
\newcommand{\Ho}{\ifmmode H_{\rm o} \else $H_{\rm o}$\fi}
\newcommand{\ho}{\ifmmode h_{\rm o} \else $h_{\rm o}$\fi}

\newcommand{\vFWHM}{\ifmmode v_{\mbox{\tiny FWHM}} \else
                    $v_{\mbox{\tiny FWHM}}$\fi}
\newcommand{\CCF}{\ifmmode F_{\it CCF} \else $F_{\it CCF}$\fi}
\newcommand{\ACF}{\ifmmode F_{\it ACF} \else $F_{\it ACF}$\fi}
\newcommand{\Halpha}{\ifmmode {\rm H}\alpha \else H$\alpha$\fi}
\newcommand{\Hbeta}{\ifmmode {\rm H}\beta \else H$\beta$\fi}
\newcommand{\Hgamma}{\ifmmode {\rm H}\gamma \else H$\gamma$\fi}
\newcommand{\Hdelta}{\ifmmode {\rm H}\delta \else H$\delta$\fi}
\newcommand{\Lya}{\ifmmode {\rm Ly}\alpha \else Ly$\alpha$\fi}
\newcommand{\Lyb}{\ifmmode {\rm Ly}\beta \else Ly$\beta$\fi}
\newcommand{\HeI}{\ifmmode {\rm He}\,{\sc i}\,\lambda5876 \else 
	          He\,{\sc i}\,$\lambda5876$\fi}
\newcommand{\HeII}{\ifmmode {\rm He}\,{\sc ii}\,\lambda4686 \else 
	           He\,{\sc ii}\,$\lambda4686$\fi}

\newcommand{\heii}{He\,{\sc ii}}

\newcommand{\ciii}{\ifmmode {\rm C}\,{\sc iii} \else C\,{\sc iii}\fi}

\newcommand{\oiii}{O\,{\sc iii}}

\newcommand{\Flamunit}{erg s$^{-1}$\,cm$^{-2}$\,\AA$^{-1}$}

\newcommand{\mbh}{$M_{\rm BH}$\ }

\newcommand{\rblr}{$R_{\rm BLR}$\ }
\newcommand{\msigma}{$M_{\rm BH}$--$\sigma_{*}$\ }
\newcommand{\mlum}{$M_{\rm BH}$--$L_{\rm Bulge}$\ }
\newcommand{\radlum}{$R_{\rm BLR}$--$L$\ }

\def\fake2{\hphantom{3}}

\shorttitle{Main Data Paper}
\shortauthors{Grier et al.}

\begin{document}

\title{Reverberation Mapping Results for Five Seyfert 1 Galaxies}

\author{C.~J.~Grier\altaffilmark{1},
B.~M.~Peterson\altaffilmark{1,2},
R.~W.~Pogge\altaffilmark{1,2},
K.~D.~Denney\altaffilmark{3},
M.~C.~Bentz\altaffilmark{4},
Paul~Martini\altaffilmark{1,2},
S.~G.~Sergeev\altaffilmark{5},
S.~Kaspi\altaffilmark{6,7},
T.~Minezaki\altaffilmark{8},
Y.~Zu\altaffilmark{1},
C.~S.~Kochanek\altaffilmark{1,2},
R.~Siverd\altaffilmark{9},
B.~Shappee\altaffilmark{1},
K.~Z.~Stanek\altaffilmark{1,2},
C.~Araya~Salvo\altaffilmark{1},
T.~G.~Beatty\altaffilmark{1},
J.~C.~Bird\altaffilmark{1},
D.~J.~Bord\altaffilmark{10},
G.~A.~Borman\altaffilmark{5,11},
X.~Che\altaffilmark{12},
C.~Chen\altaffilmark{13},
S.~A.~Cohen\altaffilmark{13},
M.~Dietrich\altaffilmark{1},
V.~T.~Doroshenko\altaffilmark{5,11,14},
T.~Drake\altaffilmark{4},
Yu.~S.~Efimov\altaffilmark{5,16},
N.~Free\altaffilmark{15},
I.~Ginsburg\altaffilmark{13},
C.~B.~Henderson\altaffilmark{1},
A.~L.~King\altaffilmark{12},
S.~Koshida\altaffilmark{8},
K.~Mogren\altaffilmark{1},
M.~Molina\altaffilmark{1},
A.~M.~Mosquera\altaffilmark{1},
S.~V.~Nazarov\altaffilmark{5,11},
D.~N.~Okhmat\altaffilmark{5,11},
O.~Pejcha\altaffilmark{1},
S.~Rafter\altaffilmark{7},
J.~C.~Shields\altaffilmark{15},
J.~Skowron\altaffilmark{1},
D.~M.~Szczygiel\altaffilmark{1},
M.~Valluri\altaffilmark{12},
and J.~L.~van~Saders\altaffilmark{1}
}

\altaffiltext{1}{Department of Astronomy, The Ohio State University,
140 W 18th Ave, Columbus, OH 43210} 
\altaffiltext{2}{Center for
Cosmology \& AstroParticle Physics, The Ohio State University, 191
West Woodruff Ave, Columbus, OH 4321, USA} 
\altaffiltext{3}{Marie Curie Fellow at the Dark
Cosmology Centre, Niels Bohr Institute, University of Copenhagen,
Juliane Maries Vej 30, DK-2100 Copenhagen, Denmark}
\altaffiltext{4}{Department of Physics and Astronomy, Georgia State
University, Astronomy Offices, One Park Place South SE, Suite 700,
Atlanta, GA 30303, USA} 
\altaffiltext{5}{Crimean Astrophysical
Observatory, P/O Nauchny Crimea 98409, Ukraine}
\altaffiltext{6}{School of Physics and Astronomy, Raymond and Beverly
Sackler Faculty of Exact Sciences, Tel Aviv University, Tel Aviv
69978, Israel} 
\altaffiltext{7}{Physics Department, Technion, Haifa
32000, Israel} 
\altaffiltext{8}{Institute of Astronomy, School of Science, University 
of Tokyo, 2-21-1, Osawa, Mitaka, 181-0015, Tokyo, Japan}
\altaffiltext{9}{Department of Physics and Astronomy, Vanderbilt University, 
5301 Stevenson Center, Nashville, TN 37235}
\altaffiltext{10}{Department of Natural Sciences, The
University of Michigan - Dearborn, 4901 Evergreen Rd, Dearborn, MI
48128} 
\altaffiltext{11}{Isaac Newton Institute of Chile, Crimean
Branch, Ukraine} 
\altaffiltext{12}{Department of Astronomy, University
of Michigan, 500 Church Street, Ann Arbor, MI 41809}
\altaffiltext{13}{Department of Physics and Astronomy, Dartmouth
College, 6127 Wilder Laboratory, Hanover, NH 03755}
\altaffiltext{14}{South Station of the Moscow MV Lomonosov State University, 
Moscow, Russia, P/O Nauchny, 98409 Crimea, Ukraine} 
\altaffiltext{15}{Department of Physics \& Astronomy, Ohio
University, Athens, OH 45701} 
\altaffiltext{16}{Deceased, 2011 October 21}

%%%%%%%%%%%%%%%%%%%%%%%%%%%%%%%%%%%%%%%%%%%%%%%%%%%%%%%%%%%%%%%%%%%%%%%%%%%%%
%%%%%%%%%%%%%%%%%%%%%%%%%%%%%%%%%%%%%%%%%%%%%%%%%%%%%%%%%%%%%%%%%%%%%%%%%%%%%
%%%%%%%%%%%%%%%%%% Main Paper %%%%%%%%%%%%%%%%%%%%%%%%%%%%%%%%%%%%%%%%%%%%%%%
%%%%%%%%%%%%%%%%%%%%%%%%%%%%%%%%%%%%%%%%%%%%%%%%%%%%%%%%%%%%%%%%%%%%%%%%%%%%%
%%%%%%%%%%%%%%%%%%%%%%%%%%%%%%%%%%%%%%%%%%%%%%%%%%%%%%%%%%%%%%%%%%%%%%%%%%%%%
\begin{abstract}
We present the results from a detailed analysis of photometric and
spectrophotometric data on five Seyfert 1 galaxies observed as a part
of a recent reverberation mapping program. The data were collected at
several observatories over a 140-day span beginning in 2010 August and
ending in 2011 January. We obtained high sampling-rate light curves
for Mrk~335, Mrk~1501, 3C\,120, Mrk~6, and PG\,2130+099, from which we
have measured the time lag between variations in the 5100\,\AA \
continuum and the \Hbeta \ broad emission line. We then used these
measurements to calculate the mass of the supermassive black hole at
the center of each of these galaxies. Our new measurements
substantially improve previous measurements of \mbh and the size of
the broad line-emitting region for four sources and add a measurement
for one new object. Our new measurements are consistent with
photoionization physics regulating the location of the broad line
region in active galactic nuclei.

\end{abstract}


\section{INTRODUCTION}

In the past two decades, several correlations have been observed
between various properties of galaxies and the masses of their central
supermassive black holes (BHs). Two of the best-studied correlations,
in both active and quiescent galaxies, are the relations between the
mass of the central black hole ($M_{\rm BH}$) and the stellar velocity
dispersion of the host bulge, commonly known as the \msigma relation
(e.g., \citealt{Ferrarese00}; \citealt{Gebhardt00a},
\citealt{Tremaine02}; \citealt{Onken04}; \citealt{Woo10}), and the
relation between \mbh and the luminosity of the host bulge, also
referred to as the \mlum relation (e.g., \citealt{Kormendy95};
\citealt{Magorrian98}; \citealt{Bentz09b}; \citealt{Gultekin09}). The
existence of these correlations suggests that there is a connection
between supermassive BH growth and galaxy evolution. If this
connection exists, simulations or theories of galaxy and BH growth
must naturally produce these observed correlations. Explanations for
the observed $M_{\rm BH}$--galaxy correlations have ranged from
hierarchical mergers and quasar feedback to self-regulated BH growth
(e.g., \citealt{Silk98}; \citealt{Dimatteo05}; \citealt{Hopkins09a}),
although there are also arguments that it is simply a consequence of
random mergers (e.g., \citealt{Peng07}; \citealt{Peng10};
\citealt{Jahnke11}).

A large sample of accurate direct \mbh measurements is crucial to
understanding this BH--galaxy connection. Because the BH sphere of
influence is much too small to be resolvable in any but the nearest
galaxies, the only direct method of measuring \mbh in distant galaxies
is reverberation mapping (\citealt{Blandford82};
\citealt{Peterson93}), which is applicable to Type 1, or broad-line,
active galactic nuclei (AGNs). Reverberation mapping relies on the
correlation between variations of the AGN continuum emission and the
subsequent response of the broad emission lines. By monitoring AGN
spectra over a period of time, one can measure the radius of the broad
line region by observing the time delay, or ``lag'', between
fluctuations in the continuum and emission-line fluxes, which is due
to light travel time between the continuum source and the
BLR. Assuming the gas is in virial motion, this BLR radius, $R_{\rm
BLR}$, can be combined with some measure of the BLR gas velocity from
the doppler-broadened emission-line widths to obtain an estimate of
$M_{\rm BH}$. To date, this method has been applied to measure BLR
radii and \mbh in nearly 50 AGNs (e.g., \citealt{Peterson04};
\citealt{Bentz09c}; \citealt{Denney10}). See \cite{Marziani12} for a
recent review on using the BLR to measure \mbh.

These measurements have confirmed the existence of a correlation
predicted by photoionization theory between the radius of the BLR and
the AGN continuum luminosity, known as the \radlum \ relation (e.g.,
\citealt{Davidson72}; \citealt{Davidson79}). This correlation allows
one to obtain both velocity and \rblr estimates from a single
calibrated spectrum, and has been used to calculate \mbh in large
samples of AGNs (e.g., \citealt{YShen08}). This can be used to
investigate the evolution of the BH mass function (e.g.,
\citealt{Greene07}; \citealt{Vestergaard08}; \citealt{Vestergaard09};
\citealt{Kelly10}), the growth of BHs compared to their hosts, the
Eddington ratios of quasars (e.g., \citealt{Kollmeier06};
\citealt{Kelly10}), and even the dependence of accretion disk sizes on
BH mass (\citealt{Morgan10}).  The existence of local correlations
between host properties and \mbh provides another means of exploring
BH populations, where BH masses can be inferred from the properties of
their hosts. However, there has recently been some discussion on the
nature of these correlations, especially the \msigma relation. In
these applications, the \msigma relation is assumed to be similar in
quiescent and active galaxies, but there are claims that many AGNs lie
below or above the \msigma relation at both the high and
low-luminosity ends (see, for example, \citealt{Dasyra07};
\citealt{Greene10}; \citealt{Mathur11}). Whether or not \mbh estimates
based on these relationships are reliable is openly
debated. Continuing to make new and improved \mbh measurements using
reverberation mapping is one way to investigate this.

Light curve quality, in terms of sampling density, duration, and
precision flux measurements, is a very important factor in
reverberation measurements. In particular, light curves that are too
short in duration or inadequately sampled can result in incorrect lag
measurements (e.g., \citealt{Perez92}; \citealt{Welsh99};
\citealt{Grier08}). Since the 1990s, our view of what constitutes
``adequately-sampled'' has changed dramatically, and we now know that
some of the early measurements need to be redone, as their sampling
rates are low enough that we have serious doubts about their
suitibility in recovering BLR radii. In a continuing effort to improve
the database of reverberation-mapped objects, we carried out a massive
reverberation mapping program at multiple institutions beginning in
2010 August and running until 2011 January. The main goals of our
program were (1) to re-observe old objects lacking well-sampled light
curves, (2) to expand the reverberation-mapped sample by observing new
objects, (3) to obtain velocity-delay maps for several of the targets,
and (4) if possible, to measure a reverberation lag in the
high-ionization \HeII \ emission line in a narrow-line Seyfert 1
galaxy (Mrk 335 in this case, with results published in
\citealt{Grier12a}). We limited our target list to galaxies with
expected time lags that were short enough to allow successful
measurements during our four-month long campaign. Our final target
list included eight objects, and we succeeded in measuring lags for
six. Two objects, NGC\,4151 and NGC 7603, were dropped due to
weather-related time losses. Here we present lag measurements for five
of the six remaining objects, while the sixth object, NGC\,7469,
presents us with a number of interesting challenges and will be
discussed in a future work. These five targets and their basic
properties are listed in Table \ref{Table:obj_info}. We will discuss
velocity-delay maps in a forthcoming study.

\section{OBSERVATIONS}
In general, we follow the observational and data reduction practices
of \cite{Denney10} for the spectroscopic observations. Our data
analysis methods follow those of \cite{Peterson04}. A brief summary
and any deviations from these methodologies are discussed below. When
needed, we adopt a cosmological model with $\Omega_{m}=0.3$,
$\Omega_{\Lambda}=0.70$, and $H_0 = 70$ km sec$^{-1}$ Mpc$^{-1}$.

\subsection{Spectroscopy}
The majority of the spectra were obtained using the MDM Observatory
1.3m McGraw-Hill telescope on Kitt Peak. We used the Boller and
Chivens CCD spectrograph to obtain spectra over the course of 120
nights from 2010 August 31 to December 28. We used the 350 mm$^{-1}$
grating to obtain a dispersion of 1.33\,\AA\,pixel$^{-1}$. We set the
grating for a central wavelength of 5150\,\AA, which resulted in
spectral coverage from roughly 4400\,\AA \ to 5850\,\AA. The slit was
oriented north-south (Position Angle=0) and set to a width of
$5''\!.0$, which resulted in a spectral resolution of 7.9 \AA. We used
an extraction window of $12''\!.75$ along the slit. We also obtained
spectra during this time period using the 2.6m Shajn telescope at the
Crimean Astrophysical Observatory (CrAO). These data were acquired
with the Nasmith spectrograph and SPEC-10 CCD. A $3''\!.0$ slit was
used at a position angle of 90$^{\circ}$, and we used an extraction
window of $13''\!.0$. Because of the large slit size used, there
should be no effect on the AGN light due to the change in position
angle between the MDM and CrAO spectra. However, this will affect the
amount of host galaxy light received through the slit. The spectral
coverage in the CrAO data was from approximately 3900 \AA \ to 6100
\AA, with a dispersion of 1.0\,\AA\,pixel$^{-1}$. Table
\ref{Table:specobs} lists the number of spectroscopic observations and
time coverage at each telescope for our sample.

\begin{figure}
\begin{center}
\epsscale{0.9}
\plotone{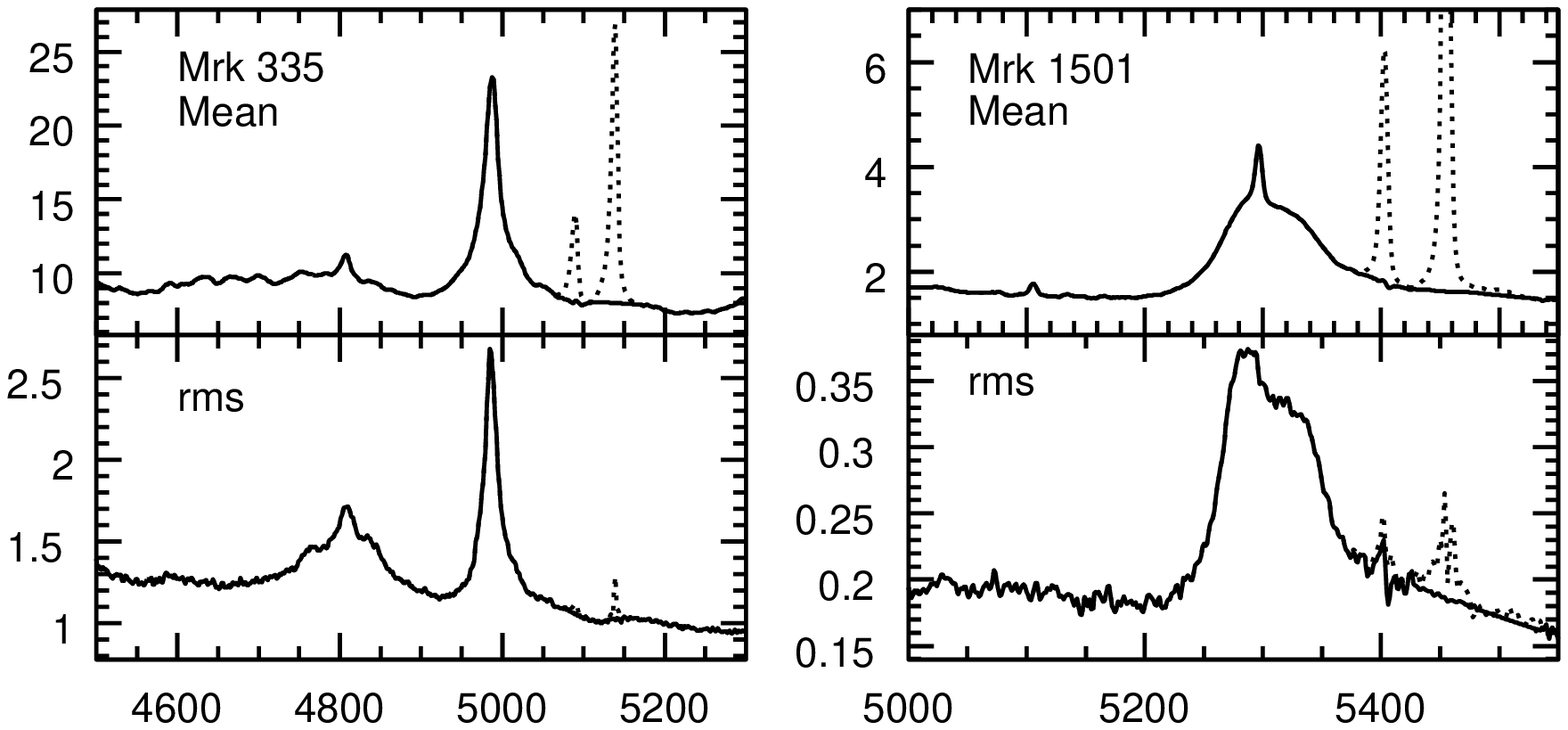}
\plotone{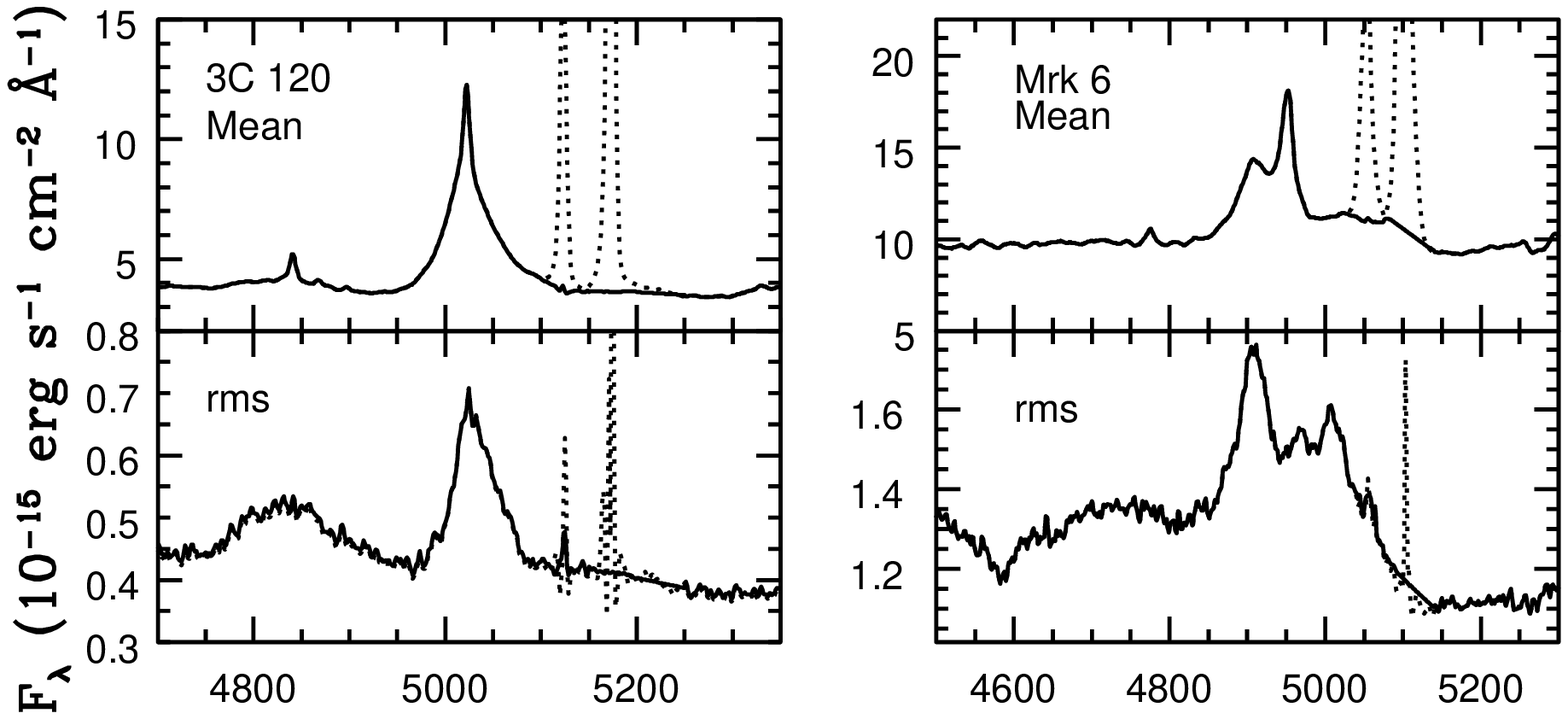}
\plotone{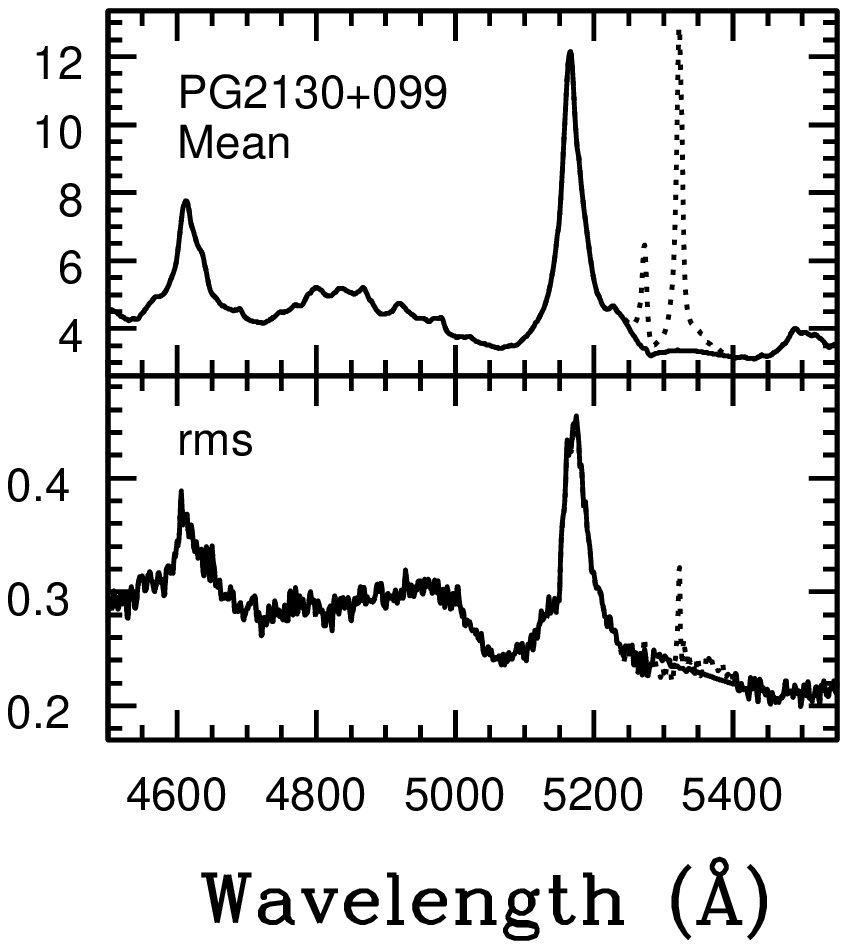}
\caption{The flux-calibrated mean and rms residual spectra of each
object. All spectra are shown in the observed frame with the flux
density in units of 10$^{-15}$ erg s$^{-1}$ cm$^{-2}$ \AA$^{-1}$. The
dotted lines show the spectra before the [O\,{\sc iii}] narrow
emission lines have been subtracted, while the solid line shows the
spectra after the subtraction. Note that we did not remove a narrow
component of the \Hbeta \ emission line.}
\label{fig:f1}
\end{center}
\end{figure}

The reduced spectra were calibrated onto an absolute flux scale by
assuming that the [O\,{\sc iii}]\,$\lambda5007$ narrow-line flux is
constant. The reference spectra for this calibration were created by
averaging spectra taken on photometric nights for each source. We
scaled these reference spectra to the absolute flux of the [O\,{\sc
iii}]\,$\lambda5007$ line for each object (listed in Column 3 of Table
\ref{Table:static}) to create an absolute flux-calibrated reference
spectrum for each object. We confirmed that the [O\,{\sc
iii}]\,$\lambda5007$ fluxes in these reference spectra agreed with
previous measurements, where available. Our new measurement of
$F$([O\,{\sc iii}]\,$\lambda5007$)~=~
3.67~$\times$~10$^{-13}$~ergs~s$^{-1}$~cm$^{-2}$ for 3C\,120 was
larger than that of \cite{Peterson98}, who measured $F$([O\,{\sc
iii}]\,$\lambda5007$) = 3.02 $\times$ 10$^{-13}$ ergs s$^{-1}$
cm$^{-2}$. Since our spectra have improved greatly in quality since
then, we adopt our new [O\,{\sc iii}]\,$\lambda5007$ flux. We did not
find a published absolute [O\,{\sc iii}]\,$\lambda5007$ flux
measurement for Mrk 1501, so for that source we adopt the flux
measured in our average spectrum of the photometric data as the
absolute [O\,{\sc iii}] flux. Using a $\chi^2$ goodness-of-fit
estimator method to minimize the flux differences between the spectra
(\citealt{vanGroningen92}), we then scaled each individual spectrum to
the reference spectrum. These procedures yield an absolute
flux-calibrated data set for each object from which to measure the
mean AGN luminosity. In some spectra, we were unable to obtain a good
fit due to changes in spectrograph focus, so we manually scaled these
spectra instead. Figure \ref{fig:f1} shows the calibrated mean and
root mean square residual (rms) spectra of our five objects based on
the calibrated MDM spectra.

\subsection{Photometry}
To supplement our spectra, we obtained $V$-band imaging observations
using the 70-cm telescope at CrAO and the 46-cm Centurion telescope
at Wise Observatory of Tel-Aviv University. The CrAO observations used
the AP7p CCD, which has 512 $\times$ 512 pixels with a 15\arcmin
$\times$15\arcmin \ field of view when mounted at prime focus. The
Wise Observatory used an STL-6303E CCD with 3072 $\times$ 2048 pixels,
with a field of view of 75\arcmin$\times$50\arcmin \ for our setup. A
summary of the photometric observations can be found in Table
\ref{Table:specobs}, including the number of observations of each
object at each telescope and their span in Heliocentric Julian Date
(HJD).

\section{LIGHT CURVES}
\subsection{Spectroscopic Light Curves}
Emission-line light curves were created for both the MDM and CrAO data
sets by fitting a linear continuum underneath the \Hbeta \ line in
each spectrum and integrating the flux above it. The continuum was
defined by two regions adjacent to the emission line, which
is defined by regions given in Table \ref{Table:int_regions}. For the
MDM data, the 5100\,\AA \ continuum light curves were created by
taking the average flux measured in the wavelength regions listed in
Table \ref{Table:int_regions}. Initial CrAO continuum and \Hbeta \
light curves were created the same way --- however, the CrAO spectra
were on a different flux scale than the MDM spectra because different
amounts of [O\,{\sc iii}] and host galaxy light enter their slits due
to changes in seeing, slit orientation, and aperture size. We assumed
there is no real variability on timescales of less than 0.5 days, so
we calibrated the CrAO light curves to the MDM light curves by
multiplying the fluxes by a constant calculated by taking the average
flux ratios between pairs of observations from the CrAO and MDM light
curves that are separated by less than 0.5 days, putting both light
curves on the same flux scale.

\subsection{Photometric Light Curves}
For the WISE imaging data, we used image subtraction to produce the
light curves using ISIS (\citealt{Alard98}; \citealt{Alard00}). We
generally follow the procedures of \cite{Shappee11}. The images are
first aligned using a program called Sexterp (Siverd 2012, in prep).
Sexterp is a replacement for ISIS' default interp.csh that relies on
SExtractor (\citealt{Bertin96}) for source identification. SExtractor
source lists are significantly more robust and improve registration
accuracy. We additionally use an upgraded interpolation utility
provided with Sexterp. This routine implements the publicly available
Bspline interpolation code of \cite{Thevenaz00} and produced better
results with our images. 

We then used ISIS to create a reference image for each field using the
20--30 images with the best seeing and lowest background counts. When
creating the reference image, ISIS convolves the images with a
spatially variable convolution kernel to transform all images to the
same point-spread function (PSF) and background level. The resulting
images are then stacked using a 3$\sigma$ rejection limit from the
median. We then used ISIS to convolve the reference image with a
kernel to match it to each individual image in the data set and
subtract each individual frame from its corresponding convolved
reference image. We then extract light curves for the nucleus of each
galaxy using ISIS to place a PSF-weighted aperture over the nucleus
and measure the residual flux. We used varying extraction apertures
for the different objects, choosing apertures large enough to account
for all AGN light but minimizing the host galaxy light included. For
the CrAO images, we used photometric fluxes based on standard aperture
photometry, which were measured within an aperture of $15''\!.0$. This
includes all of the host galaxy flux for most of our objects, and was
chosen to minimize slit losses due to variable seeing. See
\cite{Sergeev05} for more details on obtaining the CrAO photometric
fluxes.

\subsection{Combined Light Curves}
The spectroscopic continuum light curves were merged with the
photometric light curves as follows. We applied a multiplicative scale
factor as well as an additive flux adjustment to each photometric
light curve to put them all on the same scale and correct for the
differences in host galaxy starlight that enters the apertures (see
\citealt{Peterson95}). The final continuum and emission-line light
curves, scaled to our MDM light curves, are shown in Figure
\ref{fig:f2}. The continuum and \Hbeta \ fluxes are given in Tables
\ref{Table:phot} and \ref{Table:spec} and labeled according to the
observatory at which they were obtained. Final light curve statistics
are given in Table \ref{Table:lcstats}.

\section {Time-series Measurements}
Previous reverberation studies have relied on fairly simple
cross-correlation methods to measure the time delay between the
continuum and emission-line variations, $\tau$. Recently, however,
\cite{Zu11} introduced an alternative method of measuring
reverberation time lags called Stochastic Process Estimation for AGN
Reverberation (SPEAR\footnote[1]{Available at
\url{http://www.astronomy.ohio-state.edu/~yingzu/spear.html}}), and
demonstrated its ability to recover accurate time lags. We utilize
this method here. As with cross correlation, we assume all
emission-line light curves are scaled and shifted versions of the
continuum light curve. SPEAR differs from simple cross correlation
methods in two basic respects. First, SPEAR explicitly builds a model
of the light curve and transfer function and fits it to both the
continuum and the line data, maximizing the likelihood $\mathscr{L}$
of the model and then computing uncertainties using the (Bayesian)
Markov Chain Monte Carlo method. Second, as part of this process it
models the continuum light curve as an autoregressive process using a
damped random walk (DRW) model. It has long been known that AGN
continuum variability can be modeled as an autoregressive process
(\citealt{Gaskell87}) and a DRW model has been demonstrated to be a
good statistical model of quasar variability using large ($\sim 10^4$)
samples of quasar light curves (e.g., \citealt{Kelly09};
\citealt{Kozlowski10}; \citealt{MacLeod10}; \citealt{Zu12}). The
parameters of the DRW model are included in the fits and their
uncertainties, as is a simple top-hat model of the transfer function
and the light curve means (or trends if desired).

The key physical advantage of SPEAR is that it automatically includes
a self-consistent, physical model of how to interpolate in time. For
any given DRW model parameters, the stochastic process model gives a
mathematical estimate for the light curve at any time along with its
uncertainties that naturally includes all the information in both the
continuum and line light curves and their uncertainties. Since the DRW
parameters also have to be estimated from the data, we allow them to
vary as part of the overall model as well. In essence, this leads to a
lag estimate that naturally includes the uncertainties in how to
interpolate between data points, constrained by the physical
properties of the variability in the target. Because it is then a
statistical fit to the data with a set of parameters and a standard
likelihood function, it also allows the use of powerful statistical
methods like Markov Chain Monte Carlo methods to produce uncertainties
that correctly incorporate the effects of the model uncertainties on
the lag estimate.

\begin{figure}
\figurenum{4}
\begin{center}
\epsscale{0.45}
\plotone{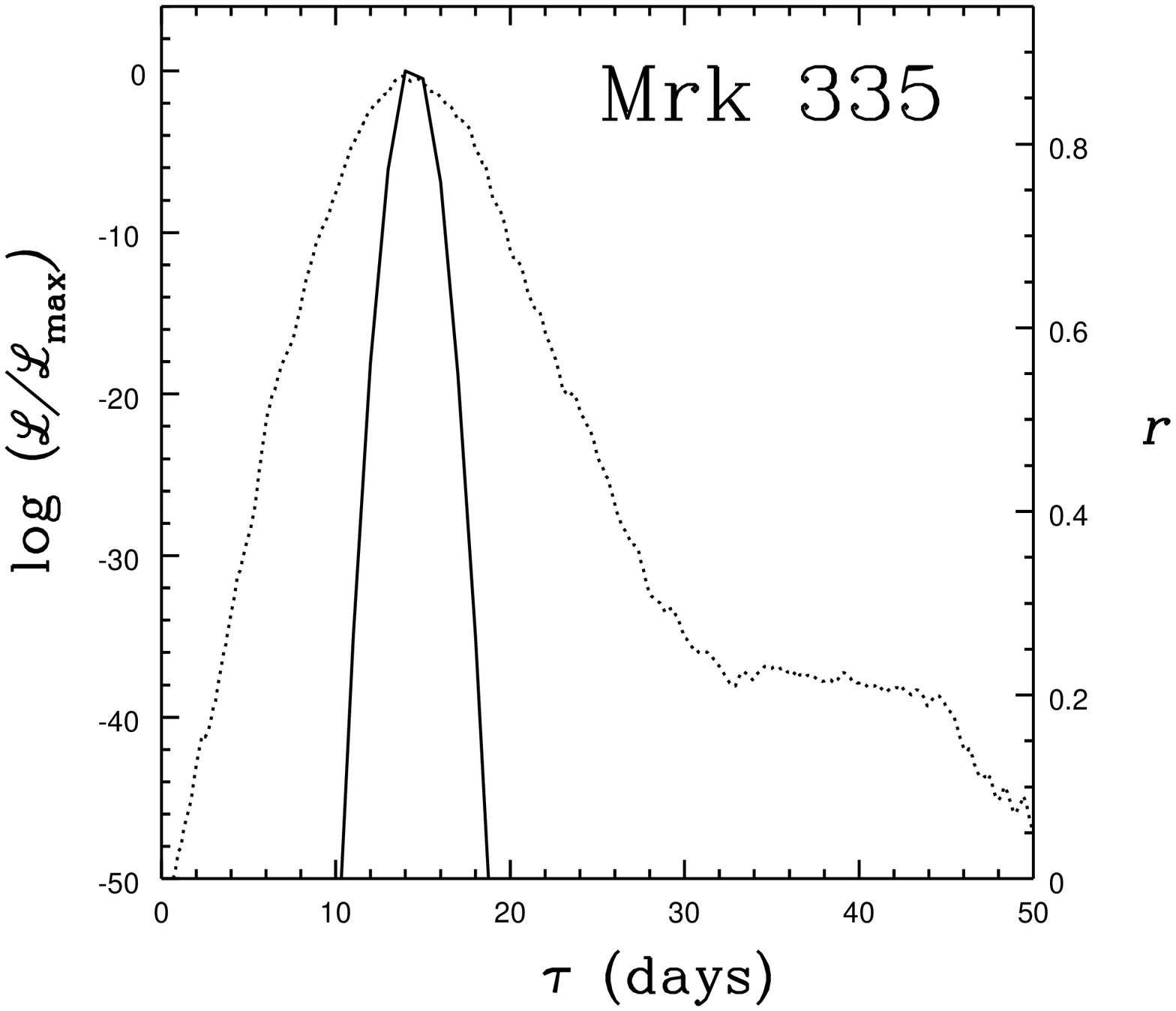}
\plotone{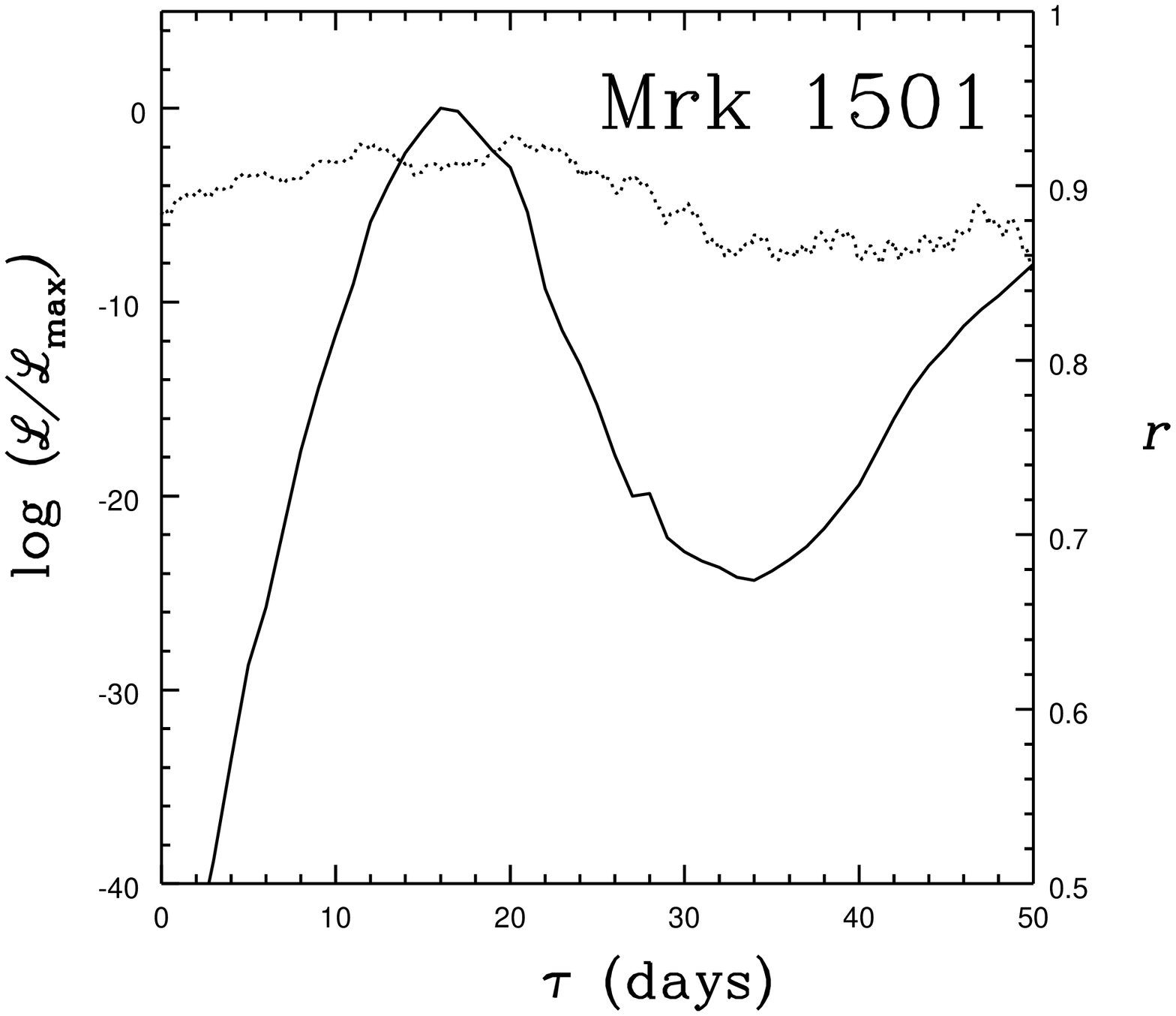}
\plotone{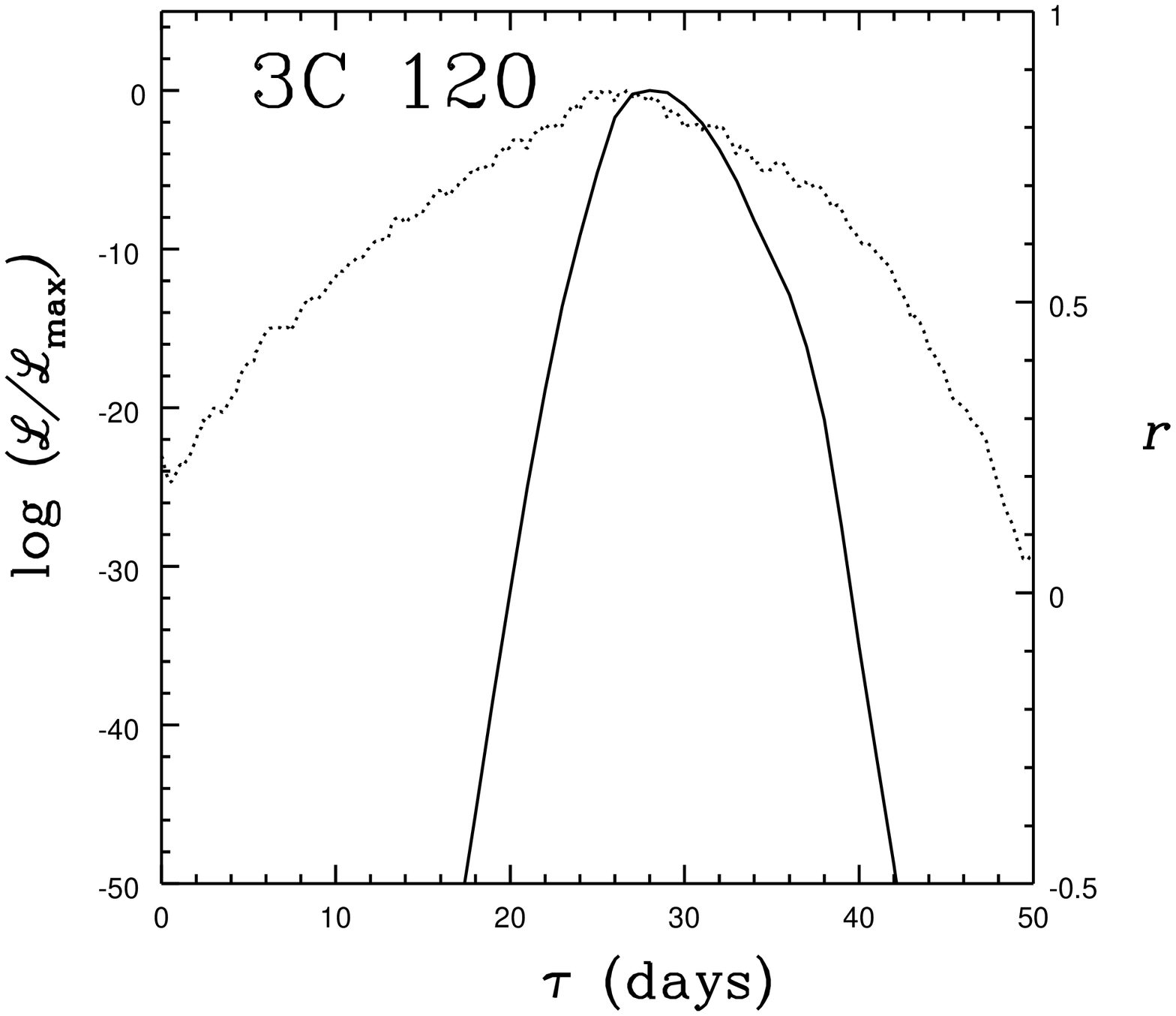}
\plotone{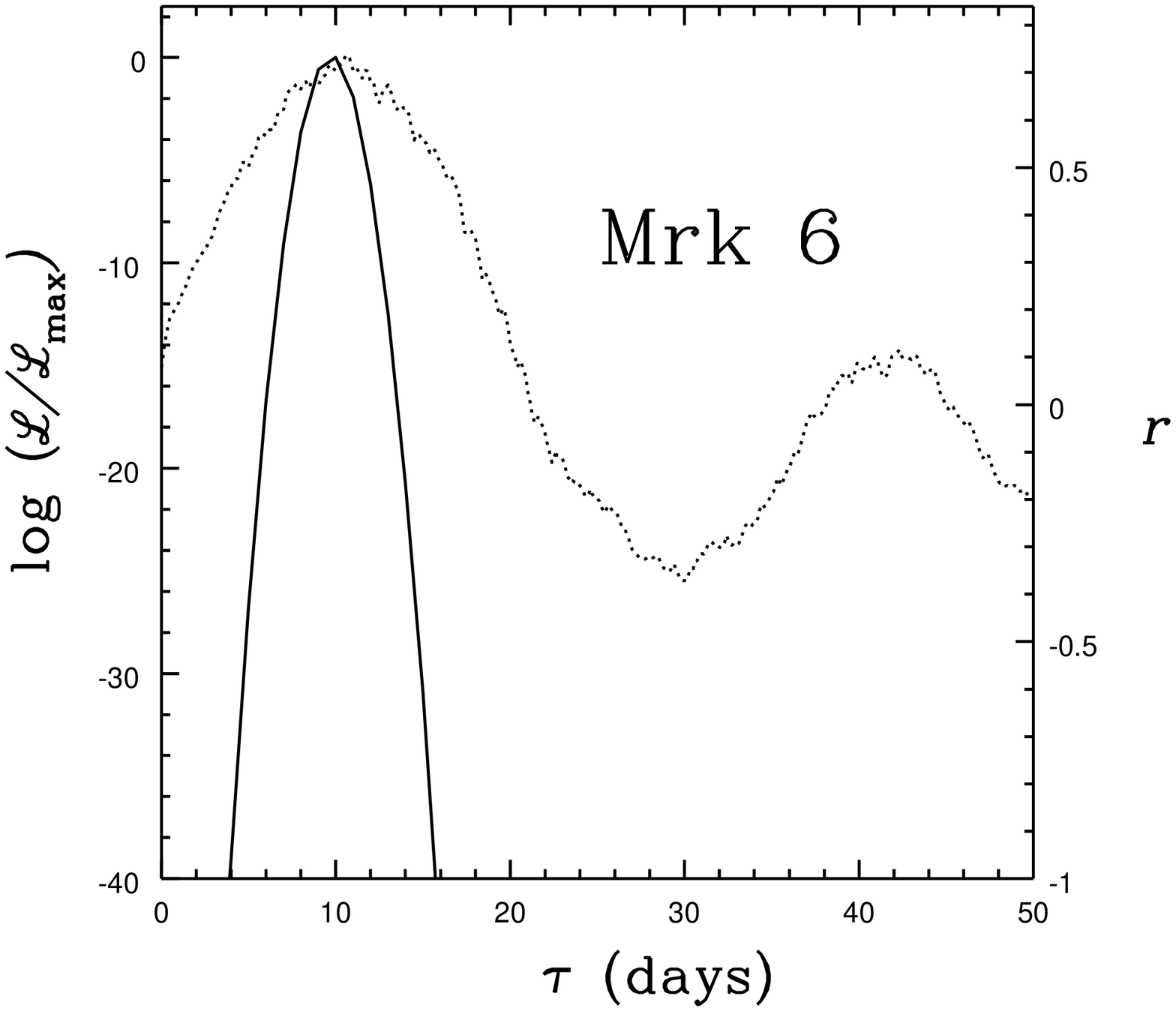}
\plotone{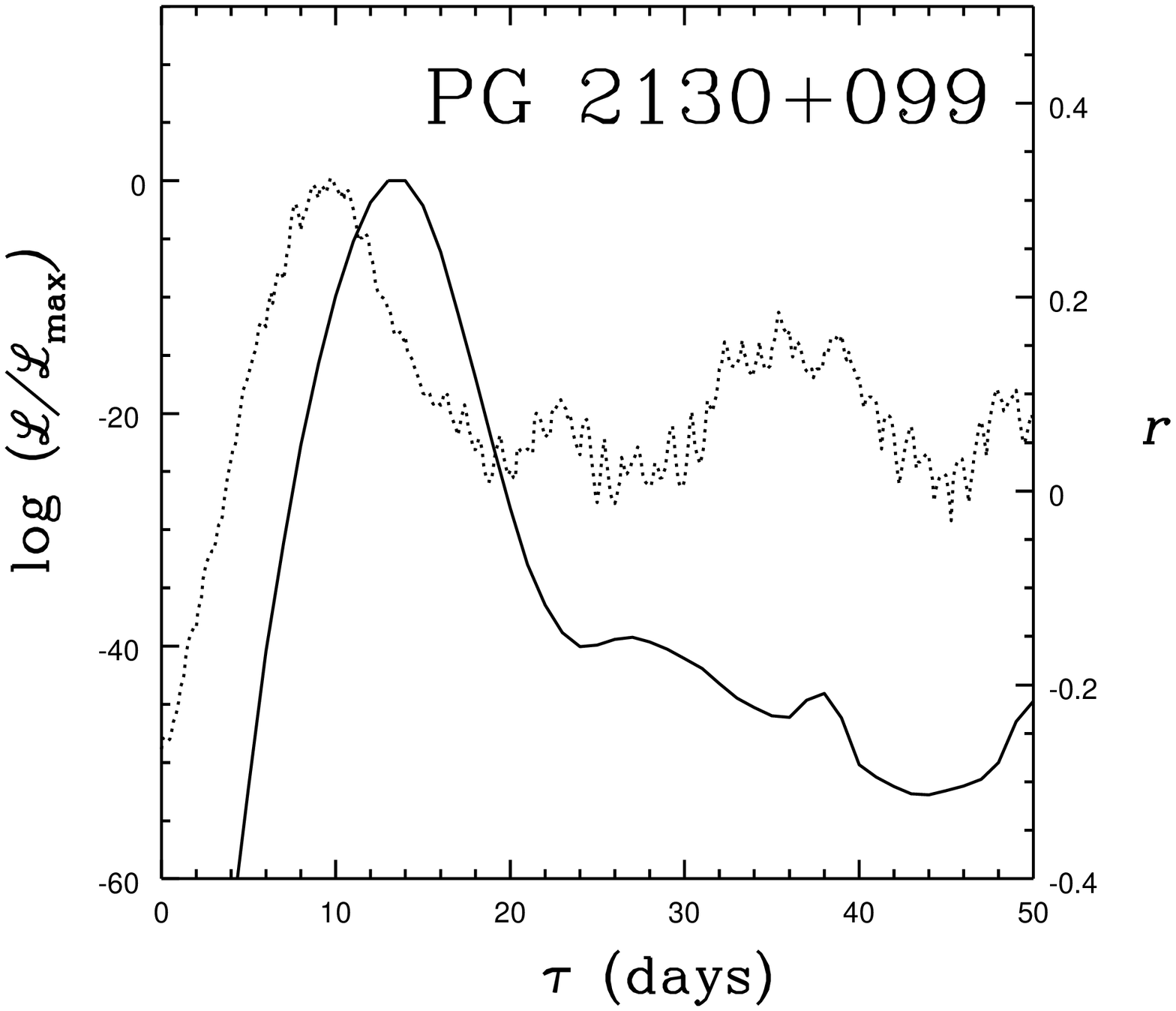}
\caption{Lag estimates. The solid black lines show the log-likelihood
functions from the SPEAR analyses, where the left axes show the SPEAR
likelihood ratios log ($\mathscr{L}$/$\mathscr{L_{\rm max}}$). The
dotted black lines show the cross correlation functions, whose $r$
values are shown on the right axes. The ranges of the y-axes were chosen
for easy comparison between the two curves.}
\label{fig:f4}
\end{center}
\end{figure}

We used SPEAR on our light curves using the code described by
\cite{Zu11} and successfully measured time lags ($\tau_{\rm SPEAR}$)
for all five objects. We list these in Table \ref{Table:lags}. The
mean and variance of the light curve models calculated by SPEAR that
are consistent with the data are shown in Figure \ref{fig:f3}. We also
show the log-likelihood functions (log ($\mathscr{L}$/$\mathscr{L_{\rm
max}})$ as a function of $\tau$) for these light curves in Figure
\ref{fig:f4}. The likelihood $\mathscr{L}$ is defined in Equation (17)
in \cite{Zu11}, and is proportional to $e^{-\chi^{2}/2}$. The best
model, corresponding to $\mathscr{L_{\rm max}}$, is associated with
the minimum $\chi^{2}$, $\chi_{min}^{2}$. Thus,
$\mathscr{L}$/$\mathscr{L_{\rm max}}$ $\propto$
$e^{(-(\chi^{2}-\chi_{min}^{2})/2)}$ and $\Delta\chi^{2}~=~-2$ ln$
(\mathscr{L}/\mathscr{L_{\rm max}})$. Therefore, Figure \ref{fig:f4}
effectively shows $\Delta\chi^{2}$ between models using each lag and
the best model.

For comparison with previous results, we also include in Table
\ref{Table:lags} the lag measurements made using the interpolation
method originally described by \cite{Gaskell86} and \cite{Gaskell87}
which was later modified by \cite{White94} and Peterson et al.\,(1998,
2004). We cross-correlate the continuum with the emission-line light
curve, calculating the value of the cross correlation coefficient $r$
at each each of many potential time lags. We show the CCFs for our
light curves in Figure \ref{fig:f4}. Uncertainties in these lags are
calculated using Monte Carlo simulations that employ the flux
randomization and random subset selection methods of
\cite{Peterson98}, as refined by \cite{Peterson04}. For each
realization, we measure the lag ($\tau_{\rm peak,CCF}$) that results
in the peak value of the cross correlation coefficient, $r_{\rm
peak}$. We also measure the lag at the centroid of the CCF ($\tau_{\rm
cent,CCF}$), calculated using points surrounding the peak with values
greater than 0.8$r_{\rm peak}$. We adopt the mean of the distribution
of delay measurements from our Monte Carlo realizations, and the
standard deviations of the same distributions are adopted as our
formal 1$\sigma$ uncertainties. In the cases of Mrk 335, Mrk 6, and
PG\,2130+099, we subtracted linear trends before performing the CCF
analysis, as there are clear secular trends in these light
curves. This did not significantly affect the measured lag values, as
can sometimes be the case. However, the resulting CCFs were cleaner,
with much more narrow and well-defined peaks when the trends were
subtracted.

\subsection {Line Width and \boldmath{\mbh}Calculations}
Assuming that the motion of the H$\beta$-emitting gas is dominated by
gravity, the relation between $M_{\rm BH}$, line width, and time delay
is
\begin{equation}
M_{\rm BH} = \frac{f c \tau \Delta V^2}{G},
\end{equation}
where $\tau$ is the measured emission-line time delay, $\Delta V$ is
the velocity dispersion of the BLR, and $f$ is a dimensionless factor
that depends on the geometry, kinematics, and orientation of the
BLR. The BLR velocity dispersion can be estimated using the observed
\Hbeta \ line width. This line width can be characterized by either
the FWHM or the line dispersion, $\sigma_{\rm line}$. To determine the
best value of the line width and its uncertainty, we use Monte Carlo
simulations similar to those used when determining the lag from the
CCF. We run 100 simulations in which we create a mean and rms residual
spectrum from a randomly chosen subset of the spectra, obtaining a
distribution of resolution-corrected line widths. We take the mean
value of $\sigma_{\rm line}$ or FWHM from these realizations and use
their standard deviation as our uncertainty. We measure $\sigma_{\rm
line}$ and FWHM in both the mean and rms residual spectra for
completeness, and report these in Table \ref{Table:mbh}. We use the
rms residual spectrum line widths to estimate $M_{\rm BH}$, as this
eliminates contamination from constant narrow line components and
isolates the broad emission components that are actually responding to
the continuum variations.

We adopt $\langle f \rangle$ = 5.5. This estimate is based on the
assumption that AGNs follow the same \msigma relationship as quiescent
galaxies (\citealt{Onken04}), and is consistent with
\cite{Woo10}. This factor allows for easy comparison with previous
results, but is about a factor of two larger than the value of
$\langle f \rangle$ computed by \cite{Graham11}. We use $\sigma_{\rm
line}$(rms) in our \mbh computation because there is at least some
evidence that it produces less biased \mbh measurements than using the
FWHM (\citealt{Peterson11}). Using $\tau_{\rm SPEAR}$ for the average
time lag, we compute the virial product $(M_{\rm vir}$ = $c \tau
\Delta V^2$/G) and \mbh for all five galaxies. The measurements are
reported in Table \ref{Table:mbh}.

%%%%%%%%%%%%%%%%%%%%%%%

\section{DISCUSSION}

\subsection{The Radius--Luminosity Relationship}
We compute the average 5100\,\AA \ luminosities of our sources,
correcting for host galaxy contamination following \cite{Bentz09a}. We
measure the observed-frame host-galaxy flux in our aperture for each
source using HST images (Table \ref{Table:static}). With these
measurements, we calculate the host-subtracted, rest-frame 5100\,\AA \
AGN luminosity for placement on the radius-luminosity
relationship. The final host-subtracted AGN luminosities are given in
Table \ref{Table:mbh}. Note that we do not currently have HST images
from which to measure the host luminosity for two of our objects, Mrk
6 and Mrk 1501. As a consequence, the luminosities listed for these
objects are the total 5100\,\AA\ luminosities rather than just that of
the AGN, and we expect them to fall to the right of the \radlum
relationship.

\begin{figure}
\figurenum{5}
\begin{center}
\epsscale{1.0}
\plotone{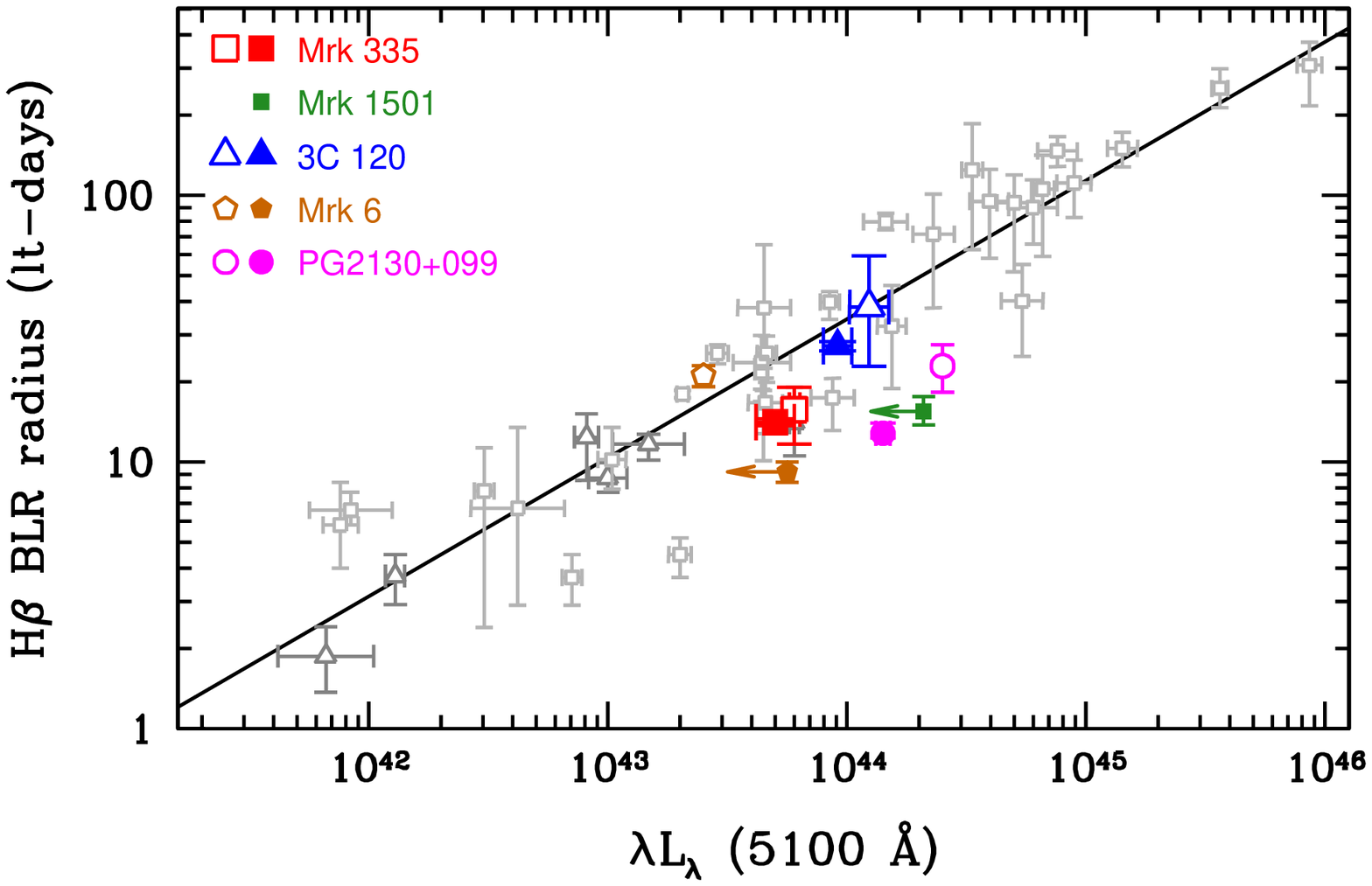}
\caption{The relationship between the BLR radius and AGN luminosity at
5100\,\AA. The most recent calibration, from \cite{Bentz09a}, is shown
by the solid line. Gray squares are from \cite{Bentz09a} and darker
gray triangles are from \cite{Denney10}. Open colored shapes show
previous measurements for our sources from \cite{Bentz09a}. The orange
open square representing Mrk 6 is from \cite{Doroshenko12}. Filled
colored shapes represent our new measurements of these objects. Each
source was given its own shape and color combination for ease of
comparison between the new and old measurements. Note that Mrk 6 and
Mrk 1501 do not have their host galaxy starlight subtracted and
therefore their continuum luminosities are shown as upper limits.}
\label{fig:f5}
\end{center}
\end{figure}

Figure \ref{fig:f5} shows the \cite{Bentz09a} \radlum relationship and
the placement of our new measurements. Previous measurements from
\cite{Bentz09a} are represented as open shapes, while our new
measurements are represented by filled shapes, varying in shape and
color by object. We have not re-fit the best-fit trend including our
new data; we leave this to a future work. Mrk 335 and 3C\,120 both
fall very close to their positions from the \cite{Bentz09a}, but we
have increased the precision of their $R_{\rm BLR}$
measurements. PG2130\,+099 continues to lie somewhat to the right of
the relation. Both Mrk 6 and Mrk 1501 also lie noticeably below the
relationship, as is expected since we were unable to subtract the host
galaxy starlight --- we therefore show these luminosity measurements
as upper limits. Host measurements for these galaxies will shift both
of them to lower luminosities and hence closer to the existing \radlum
relation.

To see where we expect Mrk 1501 and Mrk 6 to lie on the relation after
host subtraction, we examined the host galaxy light fraction in
galaxies with similar BLR sizes (i.e. similar lags) as these two
objects. Using measurements from \cite{Bentz09a}, we calculated the
average fraction of host galaxy light among galaxies with similar
lags, and used this fraction to calculate the expected host galaxy
fluxes, and hence the expected host-subtracted luminosities, in Mrk
1501 and Mrk 6. Host galaxies in objects with lags similar to Mrk 1501
contributed on average 34\% of the total luminosity, so we expect Mrk
1501 to change from log $\lambda L_{5100}$ = 44.32 $\pm$ 0.05 to
around 44.10. Host galaxies in objects with lags similar to Mrk 6
contributed on average 56\% of the total luminosity. If we applied
this to Mrk 6, the host-subtracted luminosity would then be log
$\lambda L_{5100}$ = 43.40. Both of these objects will likely continue
to lie below the current \radlum relation, but within the normal range
of scatter currently observed. However, it is important to note that
there is a very large scatter in the fraction of the luminosity
contributed by the host galaxies in general, so these numbers are used
for very rough estimations only.

\subsection{Comments on Individual Objects}

\subsubsection{Mrk 335}
Previous reverberation measurements of Mrk 335 were made by
\cite{Kassebaum97} and \cite{Peterson98} and reanalyzed by
\cite{Peterson04} and \cite{Zu11}. Previous \Hbeta \ measurements for
this object are quite good, and it was included in this study mainly
for the potential to measure the size of the high ionization component
of the BLR. Details from our study have been reported by
\cite{Grier12a}, and the data have been included in this study for
completeness. Our new measurement of \rblr = 14.1$^{+0.4}_{-0.4}$ days
is consistent with the previous measurement of \rblr =
15.3$^{+3.6}_{-2.2}$ (\citealt{Zu11}) when taking into account the
luminosity change of Mrk 335 between these two campaigns. In other
words, the position of Mrk 335 on the \radlum relationship changed
predictably given the expected photoionization slope of $R\sim
L^{1/2}$ (i.e., $\tau \sim L^{1/2}$).

\subsubsection{Mrk 1501}
No previous reverberation mapping measurements exist for Mrk 1501. We
measure $\tau$~=~15.5$^{+2.2}_{-1.9}$ days and a resulting black hole
mass of $M_{\rm BH}$~=~(1.84~$\pm$~0.27)~$\times$~10$^{8}$~\Msun. As
noted above, this object lies noticeably to the right of the \radlum
relation, which is expected since we have not yet subtracted the host
galaxy contribution to the 5100\,\AA \ luminosity due to the lack of
HST imaging data. As mentioned above, once we have corrected for host
subtraction we expect the object to lie below the relation, but still
within the normal scatter. 

\subsubsection{3C\,120}
3C\,120 was observed by \cite{Peterson98} and reanalyzed by
\cite{Peterson04}. The latter study reported $\tau_{\rm cent}$~=~
39.4$^{+22.1}_{-15.8}$~days, corresponding to \mbh~=~
5.55$^{+3.14}_{-2.25}$~$\times$~10$^{7}$~\Msun. We included 3C\,120 in
our campaign in an effort to reduce the large uncertainties in $R_{\rm
BLR}$. Our new measurement of $\tau$~=~27.2$^{+1.1}_{-1.1}$ days leads
to $M_{\rm BH}$~=~(6.7~$\pm$~0.6)~$\times$~10$^{7}$~\Msun, which is
consistent with the previous measurements, but has much smaller
uncertainties due to both better-sampled light curves and the improved
techniques of measuring lags using SPEAR. Our new measurements place
this object slightly below the \radlum relation, consistent with its
previously-measured position.

\subsubsection{Mrk 6}
Mrk 6 was observed in reverberation studies by \cite{Sergeev99},
\cite{Doroshenko03}, and \cite{Doroshenko12}, who measured \Hbeta \
time lags using cross correlation. \cite{Doroshenko12} report
$\tau_{\rm cent}$ = 21.1~$\pm$~1.9~days. This measurement was used to
calculate \mbh~=~(1.8~$\pm$~0.2)~$\times~10^{8}$ \Msun. This study
used light curves that cover a very long time period with more sparse
sampling than our campaign. Because of our dense time sampling, our
light curves are sensitive to lags as small as a day or two. We
measure a \Hbeta \ time lag of 9.2~$\pm$~0.8~days and
\mbh~=~(1.36~$\pm$~0.13)~$\times~ 10^{8}$~\Msun.

Our new $\tau$ measurement is substantially lower than the previous
measurement -- however, varying BLR sizes are expected if the
luminosity of the object changes, in accordance with the \radlum
relation. In this case, the previous study reports lower AGN
luminosity measurements than we find, and by the \radlum relation we
would also expect a smaller $\tau$ measurement in their data. However,
they measure a lag on order of twice the length of ours, so this
difference cannot be explained by a change in luminosity state. To
investigate, we ran the light curves from \cite{Doroshenko12} through
both the CCF and SPEAR analysis software, and obtain results that are
generally consistent with theirs to within errors when using cross
correlation. However, we do note that the lags we measure using SPEAR
are noticeably lower than the lags they report when we confine our
attention to their more well-sampled light curves. For example, with
their best-sampled light curves that cover the end of their observing
period, we measure $\tau$~=~11.5$^{+1.2}_{-0.8}$ days, where they
report $\tau$ = 20.4$^{4.6}_{-4.1}$ days for the same light
curves. The median spacing between observations in the
\cite{Doroshenko12} light curves is always above 10 days, which we
suspect renders their light curves insensitive to lags shorter than
this. We are confident that our measurement of $\tau$~=~9.2 days is
accurate for our data set, as the lag signal is clearly visible in our
light curves and the sampling rate is very high in both the continuum
and \Hbeta \ light curves.

Mrk 6 has a very interesting \Hbeta \ profile (see Figure
\ref{fig:f1}) that has been observed to change dramatically both in
flux and shape (\citealt{Doroshenko03}, \citealt{Sergeev99}). The rms
line profile from our study is clearly double-peaked and shows
significant blending of the \heii \ emission with the \Hbeta \
emission. To verify that our line width measurement is not affected by
the \heii \ component, we fit a second-order polynomial to the \heii \
feature in the rms spectrum and subtracted it from the total rms
spectrum. We then re-measured the line width from this new spectrum
and obtained a measurement consistent with that taken from the entire
rms spectrum. This suggests that the \heii \ blending did not affect
our measurement of $\sigma_{\rm line}$, so we adopted our original
measurement for use in the \mbh calculations. There are a variety of
physical models that can produce this double-peaked profile, many of
which we expect would show clear velocity-resolved signatures in our
data. This analysis is beyond the scope of this paper and will be
explored in detail in a future work.

\subsubsection{PG\,2130+099}
Initial reverberation results for PG\,2130+099 were first published by
\cite{Kaspi00}, who measured a value of $\tau$ on the order of 200
days and thus inferred a black hole mass of
1.4~$\times$~10$^8$~\Msun. It was a significant outlier on both the
\msigma and \radlum relations. However, PG\,2130+099 was later
re-observed and measured to have $R_{\rm
BLR}$~=~22.9$^{+4.4}_{-4.3}$~days and
\mbh~=~(3.8~$\pm$~1.5)~$\times$~$10^{7}$~\Msun \ (\citealt{Grier08}),
both of which are about an order of magnitude smaller than the
original measurements. The discrepancy was attributed to undersampled
light curves in the first measurements, as well as long-term secular
changes in the \Hbeta \ equivalent width. While the 2008 data showed a
clear reverberation signal, the amplitude of the variability in the
study was quite low and the campaign was short in duration, rendering
it insensitive to lags above 50 days, which made the light curves less
than ideal. We included this object in our study in hopes of obtaining
a better-sampled light curve sensitive to a wide range of time lags
that would yield a more definitive result. Our new measurements of
$\tau$~=~$12.8^{+1.2}_{-0.9}$~days and
\mbh~=~(4.6~$\pm$~0.4)~$\times$~$10^{7}$~\Msun \ are consistent with
those of \cite{Grier08}, but with higher precision. Note that
PG\,2130+099 is in a noticeably different position on the \radlum
relation --- it has moved nearly parallel to the relation from its
previous location, since its luminosity has also changed. Like Mrk
335, this is consistent with the expectations from photoionization
models of the BLR.

\section{SUMMARY}

We have presented reverberation measurements for five objects studied
in our 2010 observational campaign. We successfully measured the
average size of the \Hbeta-emitting region in all five objects. Four
of these measurements constitute significant improvements in precision
compared to previous measurements, and the fifth was the first
reverberation measurement for the object. We also measured the line
widths in these objects and used these to measure black hole masses,
$M_{\rm BH}$, for the sample. In all cases, our new measurements are
consistent with previous measurements, but with reduced
uncertainties. We placed our objects on the most current \radlum
relationship and find that our new measurements place our objects in
locations consistent with previous measurements when taking into
account the poorer precision of past measurements and observed mean
luminosity changes. This is consistent with the location of the BLR
being regulated by photoionization physics. We do not have host galaxy
luminosity measurements for two of our objects, and these objects lie
below the relation, as expected for objects with significant
uncorrected host galaxy contamination in their luminosities
(\citealt{Bentz09a}).

Our work also demonstrates the utility of highly-sampled light curves
in reducing uncertainties in BLR radius measurements. A large sample
of high-precision \rblr and \mbh measurements, such as the
measurements presented here, is crucial in understanding the intrinsic
scatter in the \radlum relation as well as understanding the nature of
other observed relations such as the \msigma and \mlum
relationships. We will defer discussion of these relationships to a
future contribution.

\acknowledgments We gratefully acknowledge the support of the National
Science Foundation through grant AST-1008882. BJS, CBH, and JLV are
supported by NSF Fellowships. CSK and DMS acknowledge the support of
NSF grant AST-1004756. AMM acknowledges the support of Generalitat
Valenciana, grant APOSTD/2010/030. SK is supported at the Technion by
the Kitzman Fellowship and by a grant from the Israel-Niedersachsen
collaboration program. SR is supported at Technion by the Zeff
Fellowship. SGS acknowledges the support to CrAO in the frame of the
'CosmoMicroPhysics' Target Scientific Research Complex Programme of
the National Academy of Sciences of Ukraine (2007-2012). VTD
acknowledges the support of the Russian Foundation of Research (RFBR,
project no. 09-02-01136a). The CrAO CCD cameras were purchased through
the US Civilian Research and Development for Independent States of the
Former Soviet Union (CRDF) awards UP1-2116 and UP1-2549-CR-03. This
research has been partly supported by the Grant-in-Aids of Scientific
Research (17104002, 20041003, 21018003, 21018005, 22253002, and
22540247) of the Ministry of Education, Science, Culture and Sports of
Japan. This research has made use of the NASA/IPAC Extragalactic
Database (NED) which is operated by the Jet Propulsion Laboratory,
California Institute of Technology, under contract with the National
Aeronautics and Space Administration.

%\bibliographystyle{apj} 
%\bibliography{symkate}

\clearpage

%%%%%%%%%%%%%%%%%%%%%%%%%%%%%%%%%%%%%%%%%%%%%%%%%%%%%%%%%%%%%%%%%%%%%%%%%%%%%
%%%%%%%%%%%%%%%%%%%%%%%%%%%%%%%%%%%%%%%%%%%%%%%%%%%%%%%%%%%%%%%%%%%%%%%%%%%%%
%%%%%%%%%%%%%%%%%% TABLES %%%%%%%%%%%%%%%%%%%%%%%%%%%%%%%%%%%%%%%%%%%%%%%%%%%
%%%%%%%%%%%%%%%%%%%%%%%%%%%%%%%%%%%%%%%%%%%%%%%%%%%%%%%%%%%%%%%%%%%%%%%%%%%%%
%%%%%%%%%%%%%%%%%%%%%%%%%%%%%%%%%%%%%%%%%%%%%%%%%%%%%%%%%%%%%%%%%%%%%%%%%%%%%
%1. Object info                                                            
\begin{deluxetable}{lccccccc}
\tablewidth{0pt}
\tablecaption{Object List}
\tablehead{
\colhead{ } &
\colhead{ } &
\colhead{ } &
\colhead{ } &
\colhead{ } \\
\colhead{Object} &
\colhead{RA} &
\colhead{DEC} &
\colhead{$z$} &
\colhead{A$_{B}$\tablenotemark{1}} \\
\colhead{} &
\colhead{(J2000)} &
\colhead{(J2000)} &
\colhead{} &
\colhead{(mag)}
} 
\startdata
%NGC 7469    & 23 03 15.6 & +08 52 26 & 0.0163  &  	   	\\
Mrk 335      & 00 06 19.5 & +20 12 10 & 0.0258  &  0.153  	\\
Mrk 1501     & 00 10 31.0 & +10 58 30 & 0.0893  &  0.422  	\\
3C\,120      & 04 33 11.1 & +05 21 16 & 0.0330  &  1.283  	\\
Mrk 6        & 06 52 12.2 & +74 25 37 & 0.0188  &  0.585 	\\
PG\,2130+099 & 21 32 27.8 & +10 08 19 & 0.0630  &  0.192  	
 \enddata
\label{Table:obj_info}
\tablenotetext{1}{Galactic extinctions are from \cite{Schlegel98}}. 
\end{deluxetable}

%2. Spectroscopic Info
\begin{deluxetable}{lccc|cccc}
\tablewidth{0pt}
\tablecaption{Spectroscopic and Photometric Observations}
\tablehead{
\colhead{} & \multicolumn{3}{c}{Spectroscopy} & \multicolumn{3}{c}{Photometry}  \\
\hline
\colhead{} & \colhead{} & \colhead{} & \colhead{HJD} & \colhead{} & \colhead{} & \colhead{HJD} \\
\colhead{Object} & \colhead{Observatory} & \colhead{$N_{obs}$} & \colhead{($-$2450000)} & \colhead{Observatory} & \colhead{$N_{obs}$} & \colhead{($-$2450000)}
}
\startdata
Mrk 335      & MDM  & 78 & 5440-5559   & CrAO   & 25 & 5431-5569  \\
             & CrAO & 7  & 5509-5568   & WISE   & 19 & 5511-5545  \\
Mrk 1501     & MDM  & 62 & 5440-5559   & CrAO   & 63 & 5430-5568  \\
             & CrAO & 18 & 5443-5568   & WISE   & 64 & 5433-5541  \\
3C\,120      & MDM  & 69 & 5441-5559   & CrAO   & 64 & 5430-5568  \\
             & CrAO & 15 & 5456-5569   & WISE   & 43 & 5436-5545  \\
Mrk 6        & MDM  & 75 & 5441-5562   & CrAO   & 59 & 5430-5569  \\
             & CrAO & 21 & 5443-5539   & WISE   & 50 & 5435-5545  \\
PG\,2130+099 & MDM  & 68 & 5441-5557   & CrAO   & 74 & 5430-5556  \\
             & CrAO & 20 & 5443-5539   & WISE   & 72 & 5433-5541  
\enddata			       
\label{Table:specobs}
\end{deluxetable}

%3. Static Spectral Properties. 
\begin{deluxetable}{lcccc}
\tablecolumns{4}
\tablewidth{0pt}
\tabletypesize{\small}
\tablecaption{Spectral Properties\label{tab:constspecprop}}
\tablehead{
\colhead{ }&{FWHM([\oiii]\,$\lambda 5007$)\tablenotemark{a}}&{$F$([\oiii]$\lambda$5007)}&{$F_{\rm Host}$(5100\,\AA)}\\
\colhead{Objects}&{rest frame (km s$^{-1}$)}&($10^{-13}$\,ergs s$^{-1}$\,cm$^{-2}$)&{($10^{-15} $\,\Flamunit)}\\
\colhead{(1)}&{(2)}&{(3)}&{(4)}
}
\startdata

Mrk 335      & 280     & 2.31 $\pm$ 0.10\tablenotemark{b}  	& 1.70 $\pm$ 0.16	\\
Mrk 1501     & \nodata & 1.13 $\pm$ 0.02\tablenotemark{c}       & \nodata		\\
3C\, 120     & \nodata & 3.67 $\pm$ 0.07\tablenotemark{c}  	& 0.685 $\pm$ 0.063	\\
Mrk 6        & 475     & 7.17 $\pm$ 0.12\tablenotemark{c}      &\nodata	        \\
PG\,2130+099 & 350     & 1.36 $\pm$ 0.10\tablenotemark{d}     	& 0.601 $\pm$ 0.055	

\enddata

\tablenotetext{a}{From \cite{Whittle92}.}
\tablenotetext{b}{From \cite{Peterson98}.}
\tablenotetext{c}{This work.}
\tablenotetext{d}{From \cite{Grier08}.}

\label{Table:static}
\end{deluxetable}

%4.  Integration Regions
\begin{deluxetable}{lcccc}
\tablewidth{0pt}
\tablecaption{Continuum and Emission-Line Integration Regions}
\tablehead{
\colhead{Object} &
\colhead{Continuum } & 
\colhead{\Hbeta\  Integration} \\
\colhead{} &
\colhead{Integration Region\tablenotemark{a} (\AA)} &
\colhead{Region\tablenotemark{a} (\AA)}
} 
\startdata
%NGC 7469    & 5165-5210 & 4885-5000 & \\
Mrk 335      & 5215-5240 & 4910-5100 & \\
Mrk 1501     & 5540-5560 & 5190-5540 & \\
3C\,120      & 5250-5295 & 4930-5140 & \\
Mrk 6        & 5140-5175 & 4820-5140 &  \\
PG\,2130+099 & 5420-5435 & 5085-5284 & 
\enddata
\label{Table:int_regions}
\tablenotetext{a}{All integration regions are in the observed frame.}

\end{deluxetable}

%5. Photometric/Continuum fluxes
\begin{deluxetable}{lclclclclclc}
\rotate
\tablecolumns{12}
\tablewidth{0pt}
\tabletypesize{\footnotesize}
\setlength{\tabcolsep}{0.04in}
\tablecaption{$V$-band and Continuum Fluxes\label{tab:contflux}}
\tablehead{
\multicolumn{2}{c}{Mrk 335}&\multicolumn{2}{c}{Mrk 1501}&\multicolumn{2}{c}{3C\,120}&\multicolumn{2}{c}{Mrk 6}&\multicolumn{2}{c}{PG\,2130+099}\\
\colhead{HJD\tablenotemark{a}}&{$F_{\rm con}$\tablenotemark{b}} &
\colhead{HJD\tablenotemark{a}}&{$F_{\rm con}$\tablenotemark{b}} &
\colhead{HJD\tablenotemark{a}}&{$F_{\rm con}$\tablenotemark{b}} &
\colhead{HJD\tablenotemark{a}}&{$F_{\rm con}$\tablenotemark{b}} &
\colhead{HJD\tablenotemark{a}}&{$F_{\rm con}$\tablenotemark{b}}
}
\startdata
5431.450 A  &  6.010 $\pm$ 0.070				& 5430.48 A  &  1.780 $\pm$ 0.020			 & 5430.580 A &  3.050 $\pm$ 0.090 				& 5430.548 A &  7.930 $\pm$ 0.070				& 5430.38 A &   3.560 $\pm$ 0.050 \\ 
5436.450 A  &  5.980 $\pm$ 0.090				& 5431.48 A  &  1.780 $\pm$ 0.030			 & 5432.570 A &  2.970 $\pm$ 0.050 				& 5431.526 A &  7.830 $\pm$ 0.090				& 5431.38 A &   3.670 $\pm$ 0.060 \\ 
5438.460 A  &  6.040 $\pm$ 0.080				& 5432.48 A  &  1.770 $\pm$ 0.030			 & 5433.550 A &  2.910 $\pm$ 0.060 				& 5432.517 A &  7.850 $\pm$ 0.110				& 5432.38 A &   3.550 $\pm$ 0.070 \\ 
5440.908 M  &  6.478 $\pm$ 0.091				& 5433.41 W  &  1.850 $\pm$ 0.020			 & 5436.560 A &  2.920 $\pm$ 0.050 				& 5433.491 A &  7.800 $\pm$ 0.130				& 5433.32 A &   3.560 $\pm$ 0.070 \\ 
5441.869 M  &  6.196 $\pm$ 0.087				& 5433.43 A  &  1.780 $\pm$ 0.040			 & 5436.600 W &  3.090 $\pm$ 0.010 				& 5435.574 W &  7.650 $\pm$ 0.020				& 5433.36 W &   3.380 $\pm$ 0.030 \\ 
5442.838 M  &  6.142 $\pm$ 0.087				& 5437.47 A  &  1.730 $\pm$ 0.030			 & 5437.550 A &  2.930 $\pm$ 0.040 				& 5436.482 A &  7.610 $\pm$ 0.100				& 5434.37 A &   3.520 $\pm$ 0.060 \\ 
5444.840 M  &  6.244 $\pm$ 0.088				& 5438.42 A  &  1.790 $\pm$ 0.050			 & 5437.570 W &  3.000 $\pm$ 0.020 				& 5436.586 W &  7.530 $\pm$ 0.020				& 5435.39 A &   3.510 $\pm$ 0.040 \\ 
5445.853 M  &  6.246 $\pm$ 0.088				& 5439.46 A  &  1.770 $\pm$ 0.050			 & 5439.580 A &  2.870 $\pm$ 0.080 				& 5437.487 A &  7.500 $\pm$ 0.090				& 5435.41 W &   3.480 $\pm$ 0.010 \\
\enddata    
\tablenotetext{a}{Heliocentric Julian Date ($-2450000$).}
\tablenotetext{b}{Continuum fluxes are in units of $10^{-15}$ ergs s$^{-1}$ cm$^{-2}$ \AA$^{-1}$.}
\tablenotetext{*}{Observatory Code: C=CRAO Spectroscopy, A=CRAO Photometry, W= WISE, M=MDM}
\label{Table:phot} 
\end{deluxetable}

%6. H-beta fluxes
\begin{deluxetable}{lclclclclclc}
\rotate
\tablecolumns{12}
\tablewidth{0pt}
\tabletypesize{\footnotesize}
\setlength{\tabcolsep}{0.04in}
\tablecaption{\Hbeta\ Fluxes}
\tablehead{
\multicolumn{2}{c}{Mrk 335}&\multicolumn{2}{c}{Mrk1501}&\multicolumn{2}{c}{3C\,120}&\multicolumn{2}{c}{Mrk 6}&\multicolumn{2}{c}{PG\,2130+099}\\
\colhead{HJD\tablenotemark{a}}&{$F_{\Hbeta}$\tablenotemark{b}} &
\colhead{HJD\tablenotemark{a}}&{$F_{\Hbeta}$\tablenotemark{b}} &
\colhead{HJD\tablenotemark{a}}&{$F_{\Hbeta}$\tablenotemark{b}} &
\colhead{HJD\tablenotemark{a}}&{$F_{\Hbeta}$\tablenotemark{b}} &
\colhead{HJD\tablenotemark{a}}&{$F_{\Hbeta}$\tablenotemark{b}}}

\startdata
5440.908 M &  5.323 $\pm$ 0.084  & 5440.934 M & 2.453 $\pm$ 0.067 			& 5441.957 M & 3.368 $\pm$ 0.075  			&  5442.003 M  &  6.213 $\pm$ 0.096 & 5441.725 M & 4.385 $\pm$ 0.057 \\	
5441.869 M &  5.496 $\pm$ 0.086  & 5441.893 M & 2.546 $\pm$ 0.070 			& 5443.945 M & 3.502 $\pm$ 0.078  			&  5442.959 M  &  6.129 $\pm$ 0.094 & 5442.714 M & 4.334 $\pm$ 0.056 \\
5442.838 M &  5.296 $\pm$ 0.083  & 5442.873 M & 2.537 $\pm$ 0.070 			& 5446.980 M & 3.438 $\pm$ 0.077  			&  5443.558 C  &  6.389 $\pm$ 0.229 & 5443.430 C & 4.377 $\pm$ 0.072 \\	
5444.840 M &  5.287 $\pm$ 0.083  & 5443.516 C & 2.404 $\pm$ 0.079 			& 5447.969 M & 3.373 $\pm$ 0.075  			&  5444.960 M  &  6.144 $\pm$ 0.095 & 5443.764 M & 4.300 $\pm$ 0.055 \\	
5445.853 M &  5.217 $\pm$ 0.082  & 5444.465 C & 2.459 $\pm$ 0.080 			& 5452.958 M & 3.308 $\pm$ 0.074  			&  5450.504 C  &  6.210 $\pm$ 0.223 & 5444.384 C & 4.291 $\pm$ 0.070 \\	
5447.924 M &  5.264 $\pm$ 0.083  & 5444.870 M & 2.517 $\pm$ 0.069 			& 5454.900 M & 3.309 $\pm$ 0.074  			&  5450.962 M  &  6.017 $\pm$ 0.093 & 5444.714 M & 4.391 $\pm$ 0.057 \\	
5449.974 M &  5.565 $\pm$ 0.087  & 5446.904 M & 2.508 $\pm$ 0.069 			& 5455.888 M & 3.375 $\pm$ 0.075  			&  5451.498 C  &  6.118 $\pm$ 0.219 & 5445.727 M & 4.393 $\pm$ 0.057 \\	
5451.769 M &  5.470 $\pm$ 0.086  & 5449.879 M & 2.583 $\pm$ 0.071 			& 5456.517 C & 3.319 $\pm$ 0.147  			&  5453.979 M  &  6.040 $\pm$ 0.093 & 5446.725 M & 4.303 $\pm$ 0.056 \\
\enddata 	     		     		        					 						 
		     		     		        											 
\tablenotetext{a}{Heliocentric Julian Date ($-2450000$).}
\tablenotetext{b}{\Hbeta\ flux is in units of $10^{-13}$ ergs
s$^{-1}$cm$^{-2}$.}
\tablenotetext{*}{Observatory Code: M=MDM Observatory, C=CrAO }
\label{Table:spec} 
\end{deluxetable}

%7. Light Curve Statistics
\begin{deluxetable}{lccccc|ccccc}
\rotate
\tablecolumns{12}
\tablewidth{0pt}
\tabletypesize{\small}
\tablecaption{Light Curve Statistics}
\tablehead{
\colhead{ }&\multicolumn{5}{c}{Continuum Statistics}&\multicolumn{5}{c}{\Hbeta \ Statistics}\\
\hline
\colhead{ }&\multicolumn{2}{c}{Sampling (days)}&\colhead{Mean}&{ }&{ }&\multicolumn{2}{c}{Sampling (days)}&\colhead{Mean}&{ }&{ }\\
\colhead{Objects}&{$\langle T \rangle$}&{$T_{\rm median}$}&{Flux\tablenotemark{1}}&{$F_{\rm var}$}&{$R_{\rm max}$}&{$\langle T \rangle$}&{$T_{\rm median}$}&{Flux\tablenotemark{1}}&{$F_{\rm var}$}&{$R_{\rm max}$}\\
\colhead{(1)}&{(2)}&{(3)}&{(4)}&{(5)}&{(6)}&{(7)}&{(8)}&{(9)}&{(10)}&{(11)}}

\startdata
Mrk 335       & 1.1   & 0.96  & 7.49 $\pm$ 1.01  & 0.13  & 1.57 $\pm$ 0.04   & 1.5\fake2    	& 1.00   & 5.74 $\pm$ 0.55    & 0.09  & 1.35 $\pm$ 0.03  \\
Mrk 1501      & 0.66  & 0.48  & 1.49 $\pm$ 0.16  & 0.11  & 1.48 $\pm$ 0.04   & 1.55   		& 0.99   & 2.16 $\pm$ 0.23    & 0.10  & 1.44 $\pm$ 0.05  \\
3C\,120       & 0.72  & 0.53  & 3.37 $\pm$ 0.38  & 0.11  & 1.49 $\pm$ 0.07   & 1.5\fake2    	& 0.99   & 3.75 $\pm$ 0.24    & 0.06  & 1.28 $\pm$ 0.04  \\
Mrk 6         & 0.84  & 0.58  & 8.93 $\pm$ 1.14  & 0.13  & 1.65 $\pm$ 0.04   & 1.2\fake2    	& 0.99   & 7.00 $\pm$ 0.69    & 0.10  & 1.42 $\pm$ 0.04 \\
\enddata										                
\tablenotetext{1}{Continuum and emission-line fluxes are given in $10^{-15}$ \Flamunit \ and $10^{-13}$ \ergscm, respectively, and have not been corrected 
for host galaxy contamination.}					
\tablenotetext{*} {Column (1) lists the object, Columns (2) and (3) list the average and median time spacing between continuum observations, 
respectively. Column (4) gives the mean flux of the continuum in the observed frame. Column (5) gives the excess variance, defined by
\begin{equation}
F_{\rm var} = \frac{\sqrt{\sigma^{2}-\delta{^2}}}{\langle f\rangle}
\end{equation}
where $\sigma^{2}$ is the flux variance of the observations, $\delta^{2}$ is the mean square uncertainty, and $\langle f \rangle$ 
is the mean observed flux (\citealt{Rodriguezpascual97}). Column (6) is the ratio of the maximum to
minimum flux in each light curve. Columns 7-11 are the
same quantities but computed using the merged \Hbeta \ light
curves rather than the continuum light curves.}	   
\label{Table:lcstats}									   
\end{deluxetable}

%8. Time Series Results
\begin{deluxetable}{lccc} 
\tablewidth{0pt} 
\tablecaption{Rest-frame \Hbeta \ Lag Measurements}                                                                                       
\tablehead{ 
\colhead{Object} & 
\colhead{$\tau_{\rm SPEAR}$} &
\colhead{$\tau_{\rm cent,CCF}$}   &
\colhead{$\tau_{\rm peak,CCF}$}  \\ 
\colhead{} & 
\colhead{(days)} & 
\colhead{(days)} & 
\colhead{(days)} \\  
\colhead{(1)} & 
\colhead{(2)} & 
\colhead{(3 } & 
\colhead{(4)}  
} 
\startdata
Mrk 335       & 14.1$^{+0.4}_{-0.4}$  & 14.3 $\pm$ 0.7  	&  14.0 $\pm$ 0.9   \\
Mrk 1501      & 15.5$^{+2.2}_{-1.8}$  & 12.6 $\pm$ 3.9  	&  13.8 $\pm$ 5.4   \\  
3C\,120       & 27.2$^{+1.1}_{-1.1}$  & 25.9 $\pm$ 2.3  	&  25.6 $\pm$ 2.4   \\   
Mrk 6         & \fake29.2$^{+0.8}_{-0.8}$   & 10.1 $\pm$ 1.1   	&  10.2 $\pm$ 1.2   \\
PG\,2130+099  & 12.8$^{+1.2}_{-0.9}$  & 9.6 $\pm$ 1.2  	&  9.7 $\pm$ 1.3    
\enddata
\label{Table:lags}
\end{deluxetable}

%9. Line Widths & MBH 
\begin{deluxetable}{lcccccccc} 
\rotate
\tablewidth{0pt} 
\tablecaption{Mean and RMS Line Widths, Virial Masses and Luminosities}                                                                                       
\tablehead{ 
\colhead{Object} & 
\colhead{$\sigma_{\rm line}$(mean)} &
\colhead{FWHM(mean)} &
\colhead{$\sigma_{\rm line}$(rms)}  &
\colhead{FWHM(rms)}  &
\colhead{$M_{\rm vir}$}  &
\colhead{\mbh}  &
\colhead{log $\lambda L_{5100}$} \\ 
\colhead{} & 
\colhead{(\kms)} & 
\colhead{(\kms)} & 
\colhead{(\kms)} & 
\colhead{(\kms)} & 
\colhead{($\times$10$^{6}$\Msun)} & 
\colhead{($\times$10$^{6}$\Msun)} & 
\colhead{(erg s$^{-1}$)}\\  
\colhead{(1)} & 
\colhead{(2)} & 
\colhead{(3)} & 
\colhead{(4)} & 
\colhead{(5)} & 
\colhead{(6)} &
\colhead{(7)}&
\colhead{(8)} 
} 
\startdata
Mrk 335       & 1663 $\pm$ 6\fake2   & 1273 $\pm$ 3\fake2     &   1293 $\pm$ 64  & 1025 $\pm$ 35\fake2    & \fake24.6 $\pm$ 0.5   	& \fake225 $\pm$ 3\fake2   & 43.70 $\pm$ 0.08	   \\	
Mrk 1501      & 3106 $\pm$ 15  	     & 3494 $\pm$ 35          &   3321 $\pm$ 107 & 5054 $\pm$ 145  	  & 33.4 $\pm$ 4.9   		& 184 $\pm$ 27  	   & 44.32 $\pm$ 0.05      \\  
3C\,120       & 1687 $\pm$ 4\fake2  & 1430 $\pm$ 16          &   1514 $\pm$ 65  & 2539 $\pm$ 466         & 12.2 $\pm$ 1.2 		& \fake267 $\pm$ 6\fake2   & 43.96 $\pm$ 0.06      \\   
Mrk 6         & 4006 $\pm$ 6\fake2   & 2619 $\pm$ 24          &   3714 $\pm$ 68  & 9744 $\pm$ 370  	  & 24.8 $\pm$ 2.3  		& 136 $\pm$ 12  	   & 43.75 $\pm$ 0.06      \\   
PG\,2130+099  & 1760 $\pm$ 2\fake2   & 1781 $\pm$ 5\fake2     &   1825 $\pm$ 65  & 2097 $\pm$ 102  	  & \fake28.3 $\pm$ 0.7   	& \fake246 $\pm$ 4\fake2   & 44.15 $\pm$ 0.03            
\enddata
\label{Table:mbh}
\end{deluxetable}

\clearpage
%%%%%%%%%%%%%%%%%%%%%%%%%%%%%%%%%%%%%%%%%%%%%%%%%%%%%%%%%%%%%%%%%%%%%%%%%%
%%%%%%%%%%%%%%%%%%%%%%%%%%%%%%%%%%%%%%%%%%%%%%%%%%%%%%%%%%%%%%%%%%%%%%%%%%%%%
%%%%%%%%%%%%%%%%%%%%%%%%%%%%%%%%%%%%%%%%%%%%%%%%%%%%%%%%%%%%%%%%%%%%%%%%%%%%%
%%%%%%%%%%%%%%%%%% FIGURES %%%%%%%%%%%%%%%%%%%%%%%%%%%%%%%%%%%%%%%%%%%%%%%%%%
%%%%%%%%%%%%%%%%%%%%%%%%%%%%%%%%%%%%%%%%%%%%%%%%%%%%%%%%%%%%%%%%%%%%%%%%%%%%%
%%%%%%%%%%%%%%%%%%%%%%%%%%%%%%%%%%%%%%%%%%%%%%%%%%%%%%%%%%%%%%%%%%%%%%%%%%%%%

%%%%%%%%%%%%%%%%Mean and rms spectra %%%%%%%%%%%%%%%%%%%%%%%%%%%%%

%%%%%%%%%%%%%LIGHT CURVES %%%%%%%%%%%%%%%%%%%%%%%%%%%%%%%%%%%%%%%
\begin{figure}
\figurenum{2}
\epsscale{1.0}
\plotone{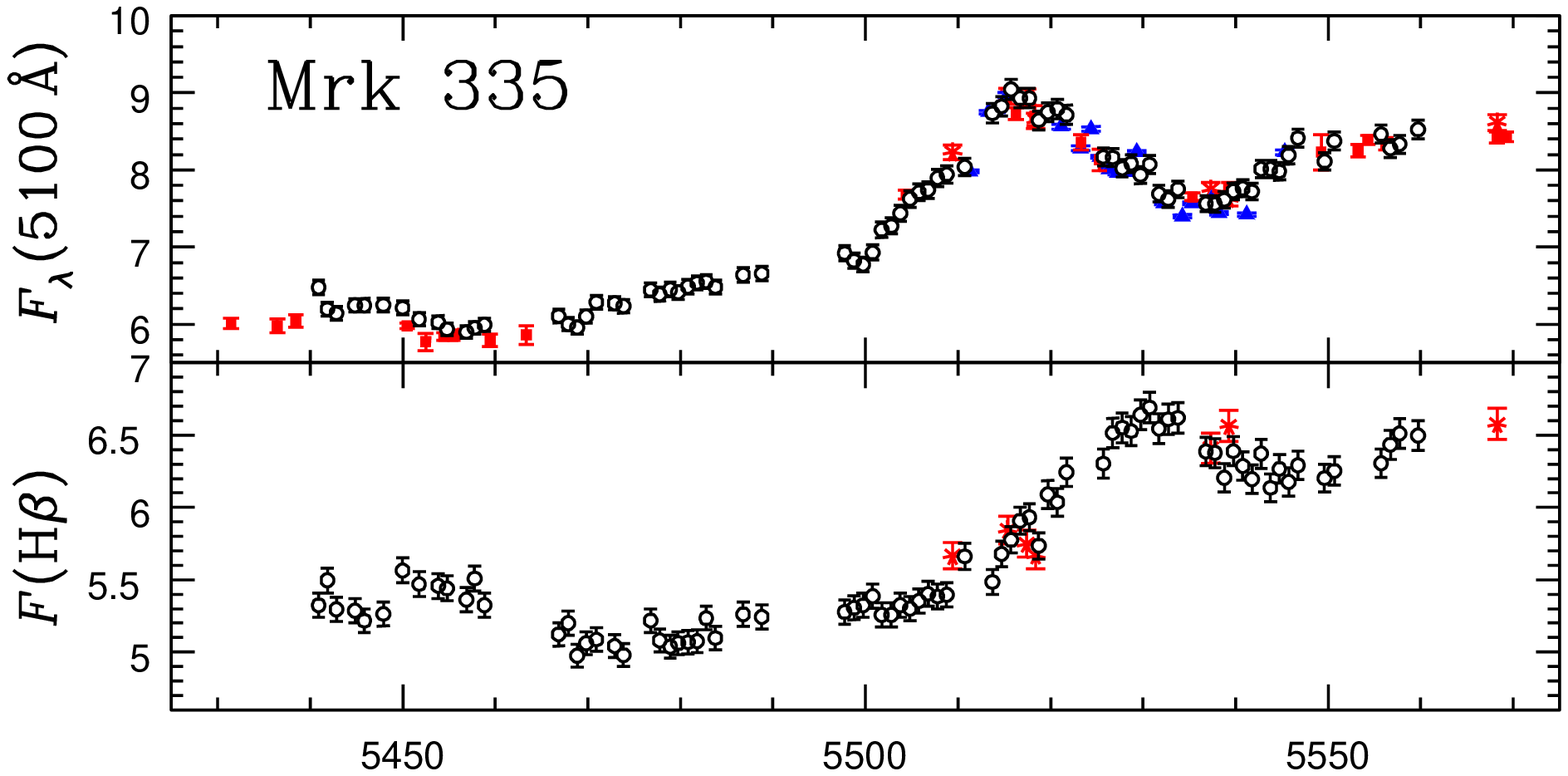}
\plotone{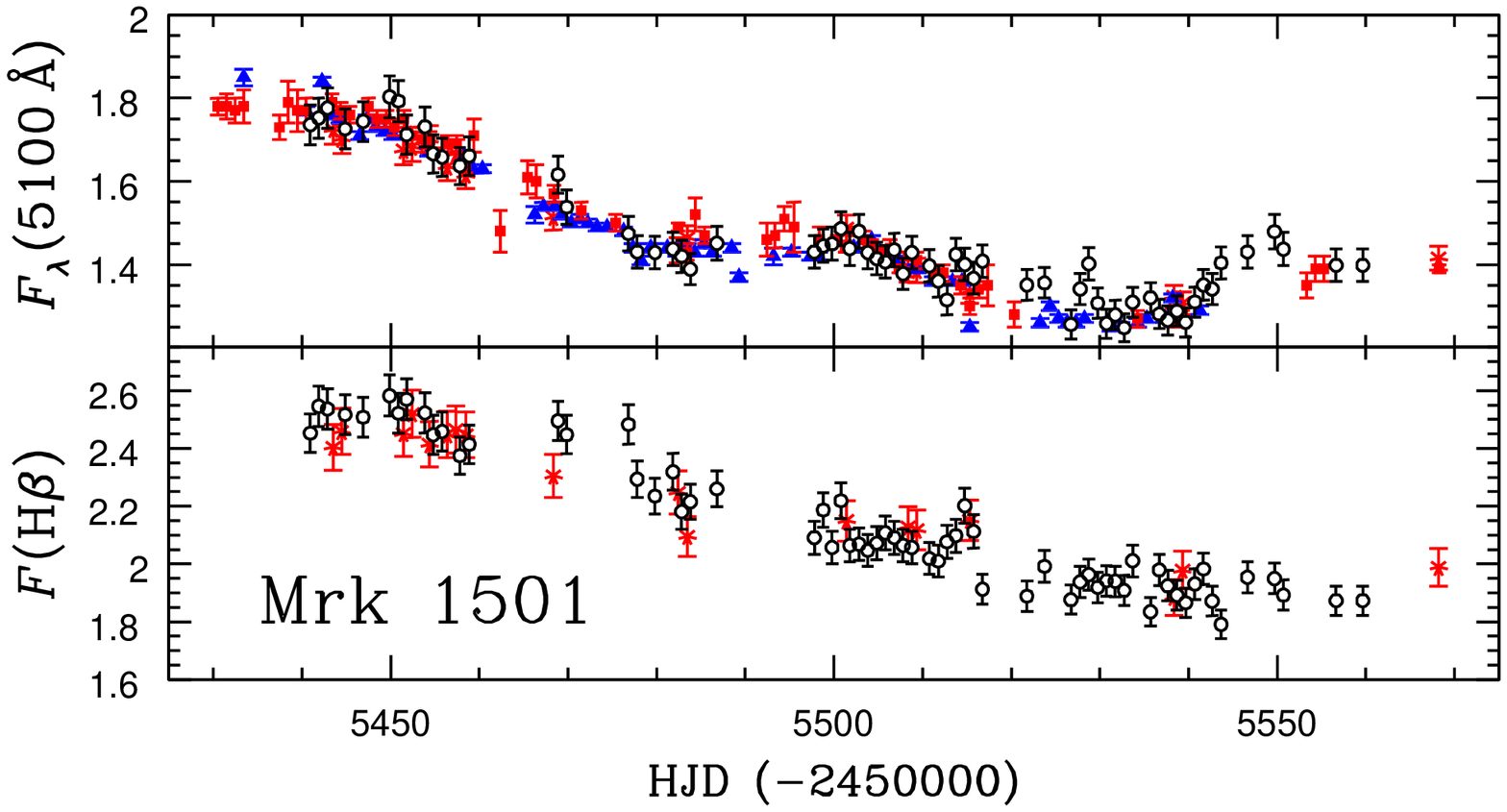}

\caption{Complete light curves for the five objects observed during
our campaign. For each object, the top panel shows the 5100\,\AA \ flux
in units of $10^{-15}$ \ergscm\AA$^{-1}$ and the bottom panel shows
the integrated H$\beta\lambda4861$ flux in units of $10^{-13}$
\ergscm. Open black circles denote the observations from MDM
Observatory and red asterisks represent spectra taken at CrAO. Closed
red squares show the photometric observations from CrAO, and closed
blue triangles represent photometric observations from the WISE
Observatory. }

\label{fig:f2}
\end{figure}

\begin{figure}\ContinuedFloat
\figurenum{2}
\ContinuedFloat
\epsscale{1.0}
\plotone{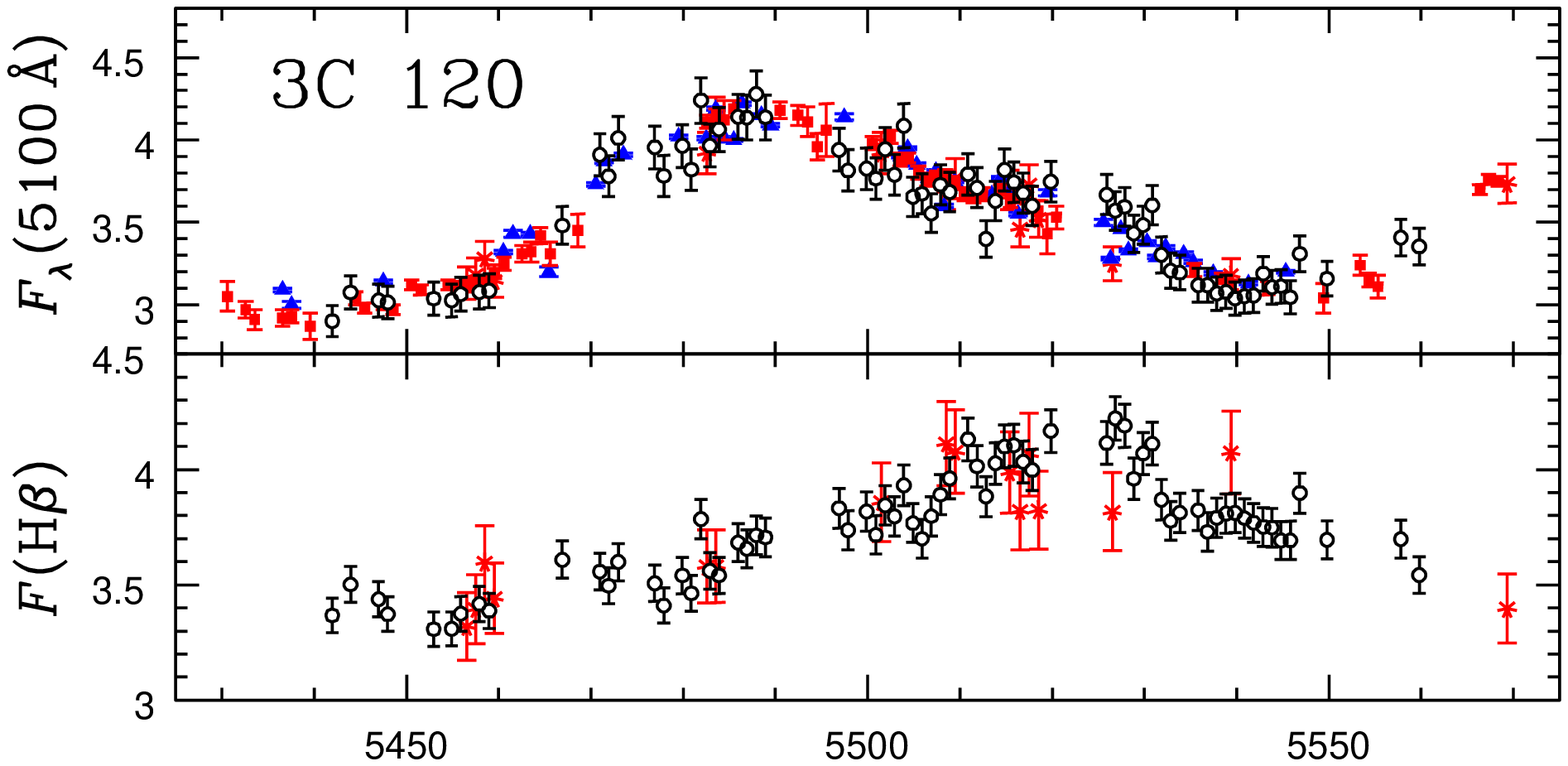}
\plotone{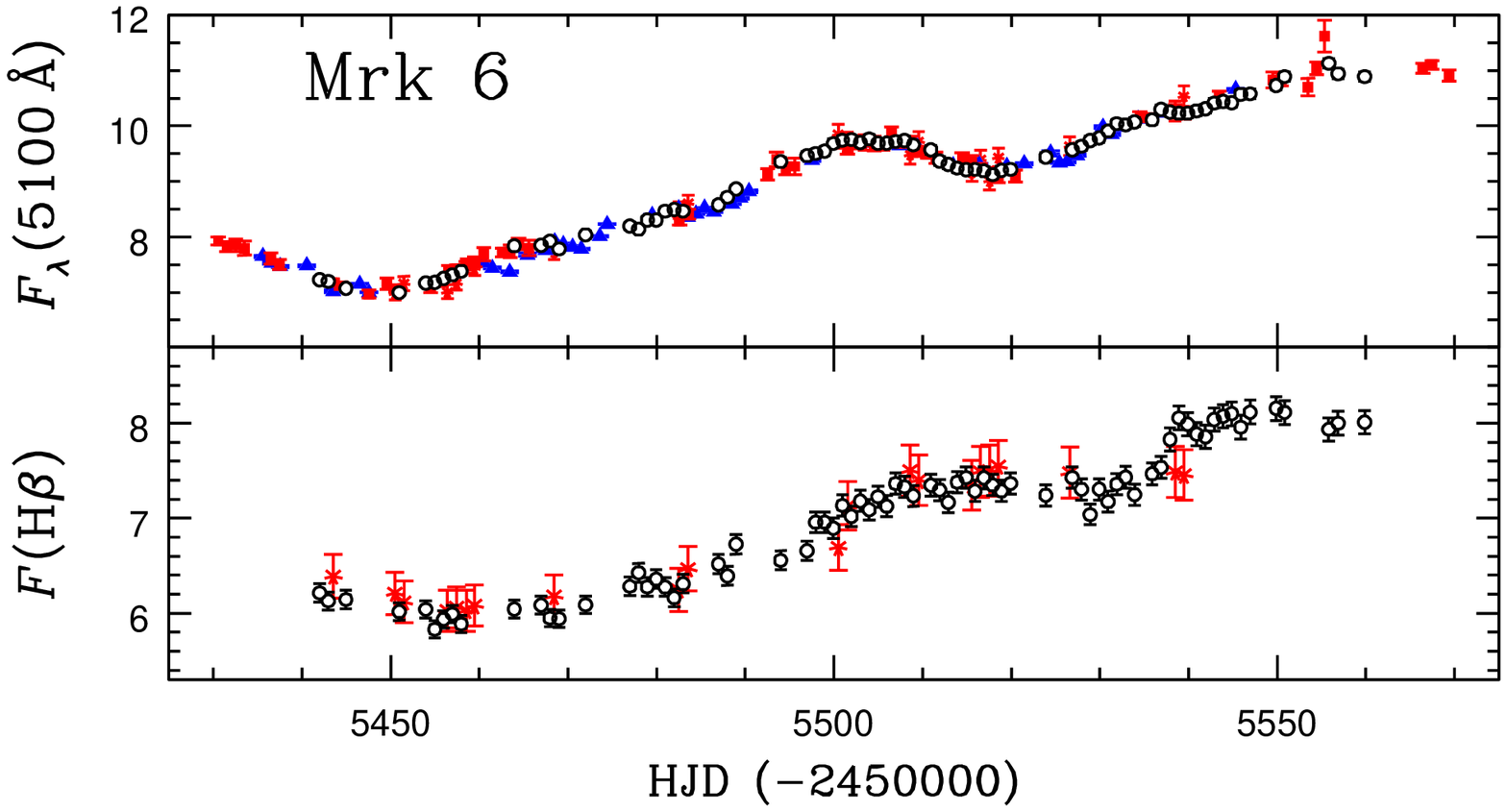}

\caption{{\it Continued.}}

\end{figure}

\begin{figure}\ContinuedFloat
\figurenum{2}
\ContinuedFloat
\epsscale{1.0}
%\plotone{figures/fig2f.eps}
\plotone{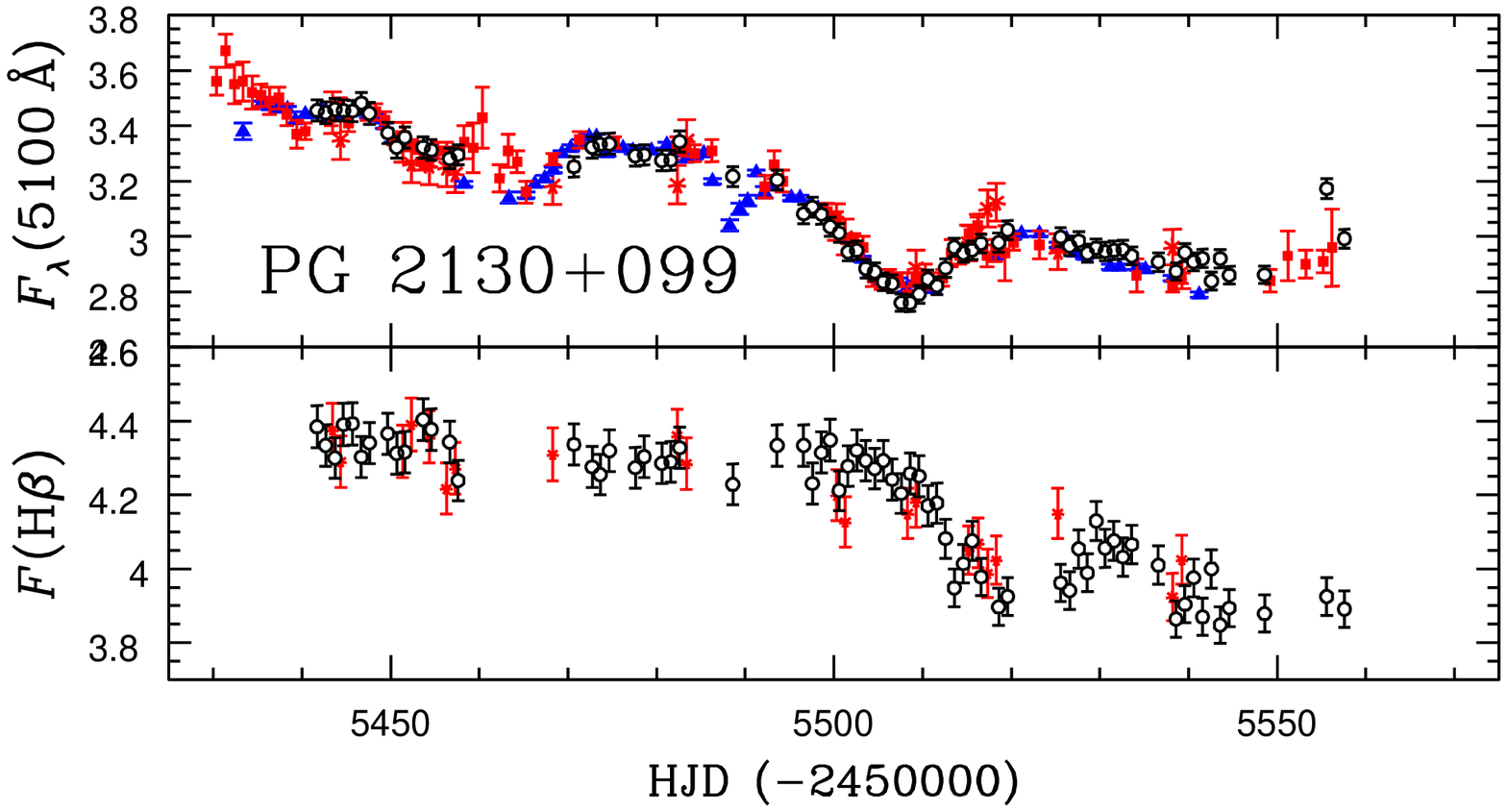}

\caption{{\it Continued.}}

\end{figure}

%%%%%%%%%%%PREDICTED LIGHT CURVES FROM SPEAR %%%%%%%%%%%%%%%%%%%%%%
\begin{figure}
\figurenum{3}
\epsscale{1.0}
\plotone{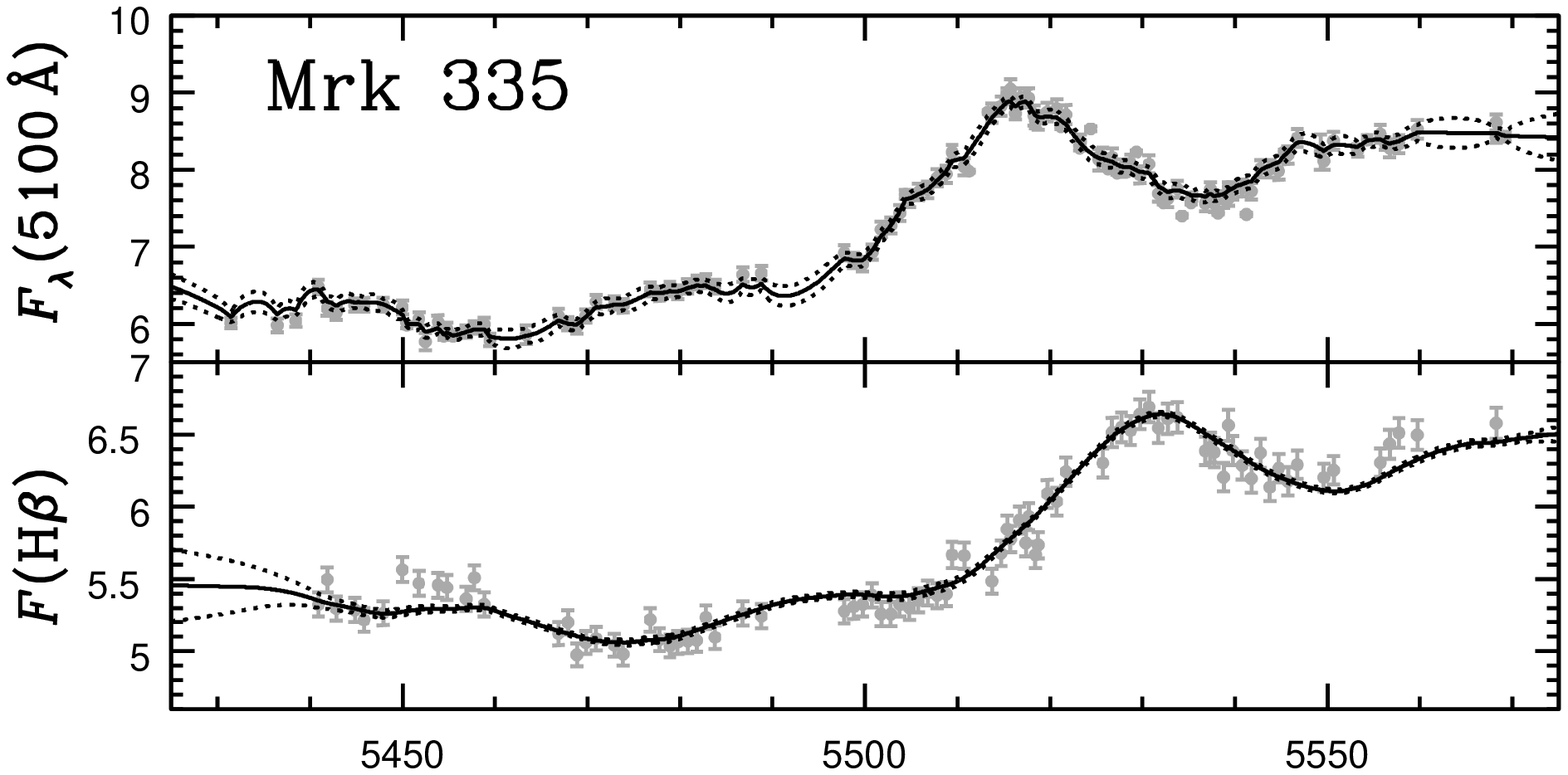}
\plotone{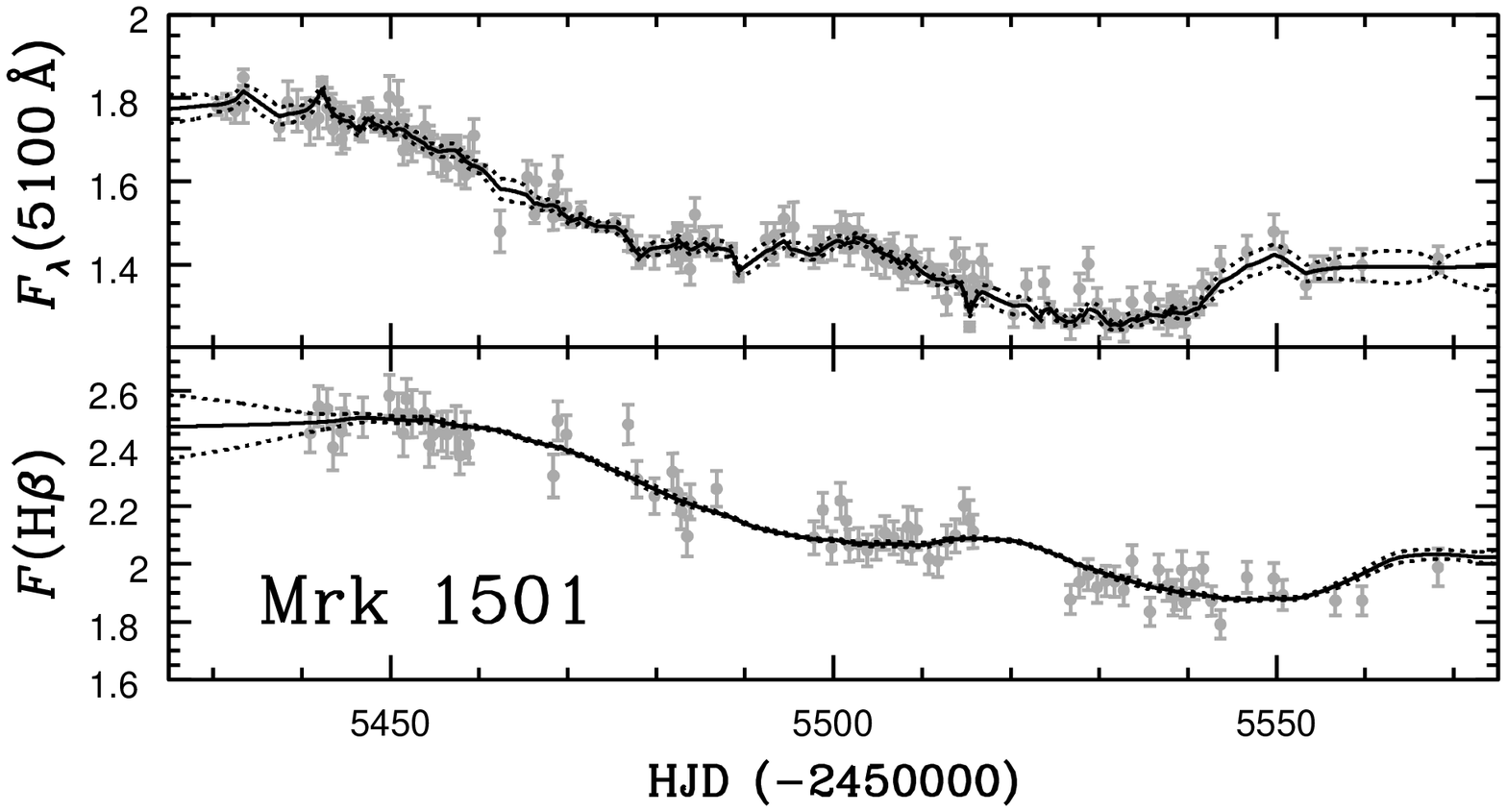}

\caption{The mean of the predicted light curves and their dispersionas
as estimated by the best-fit SPEAR model. For each object, the top
panel shows the continuum light curve, and the bottom panel shows
the \Hbeta \ light curve, both in the same units as Figure 2. The
gray points show the merged light curves used in the model, and the
solid line shows the mean of the SPEAR light curve models fit to the
data. Dotted black lines show the standard deviation of values about
the mean (see \citealt{Zu11}).}
 
\label{fig:f3}
\end{figure}

\begin{figure}\ContinuedFloat
\figurenum{3}
\ContinuedFloat
\epsscale{1.0}
\plotone{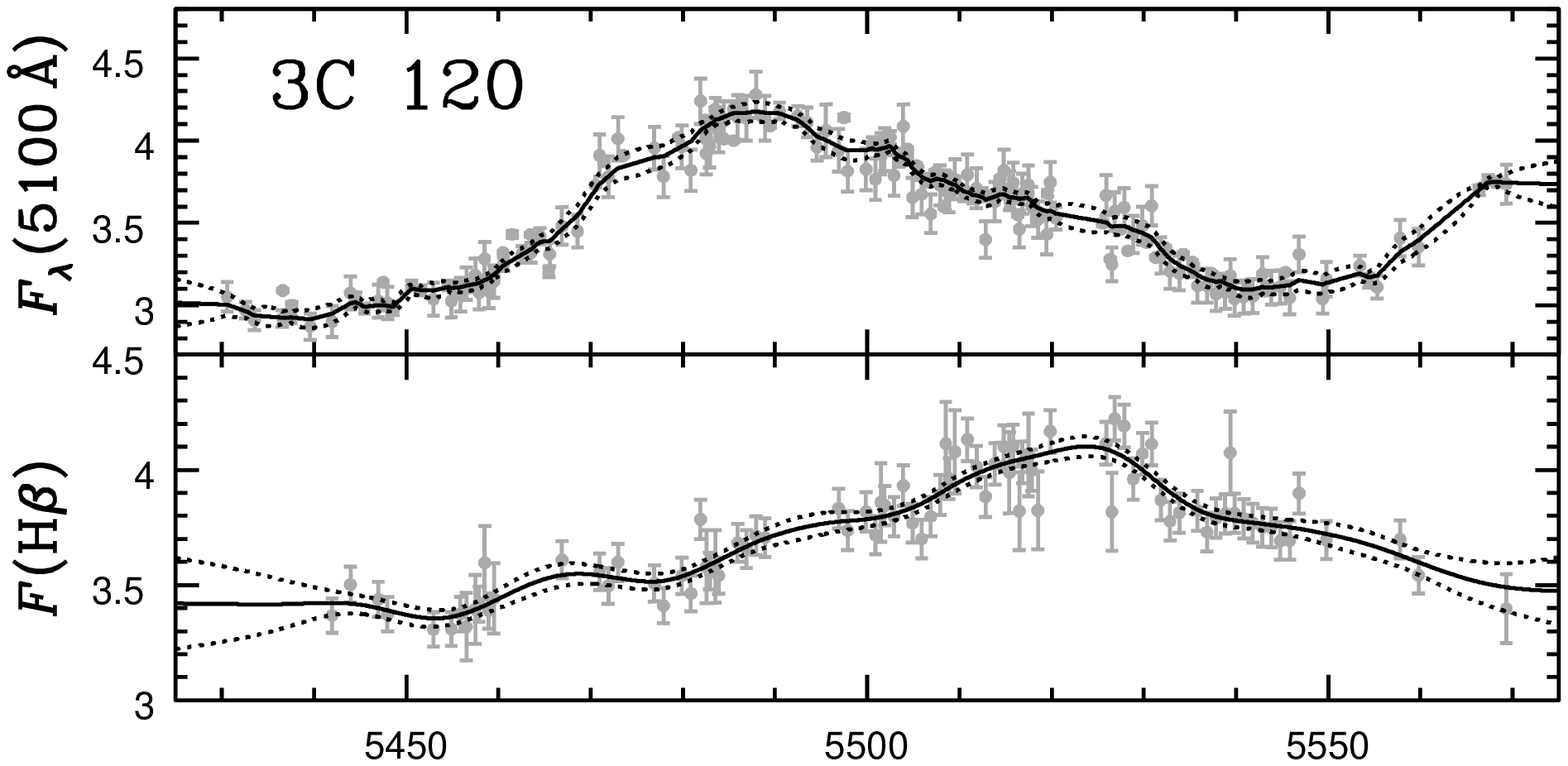}
\plotone{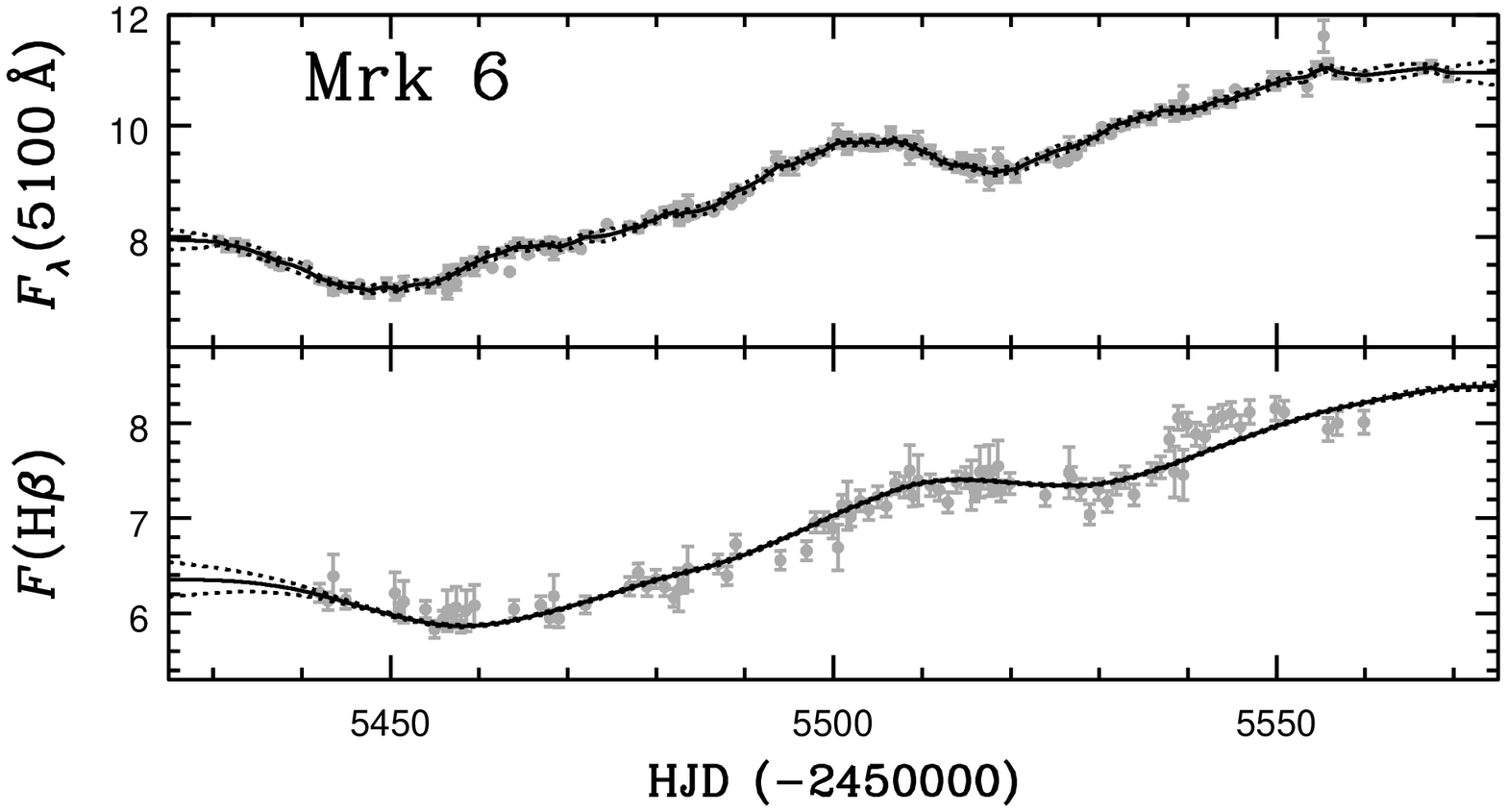}

\caption{{\it Continued.}}

\end{figure}

\begin{figure}\ContinuedFloat
\figurenum{3}
\ContinuedFloat
\epsscale{1.0}
\plotone{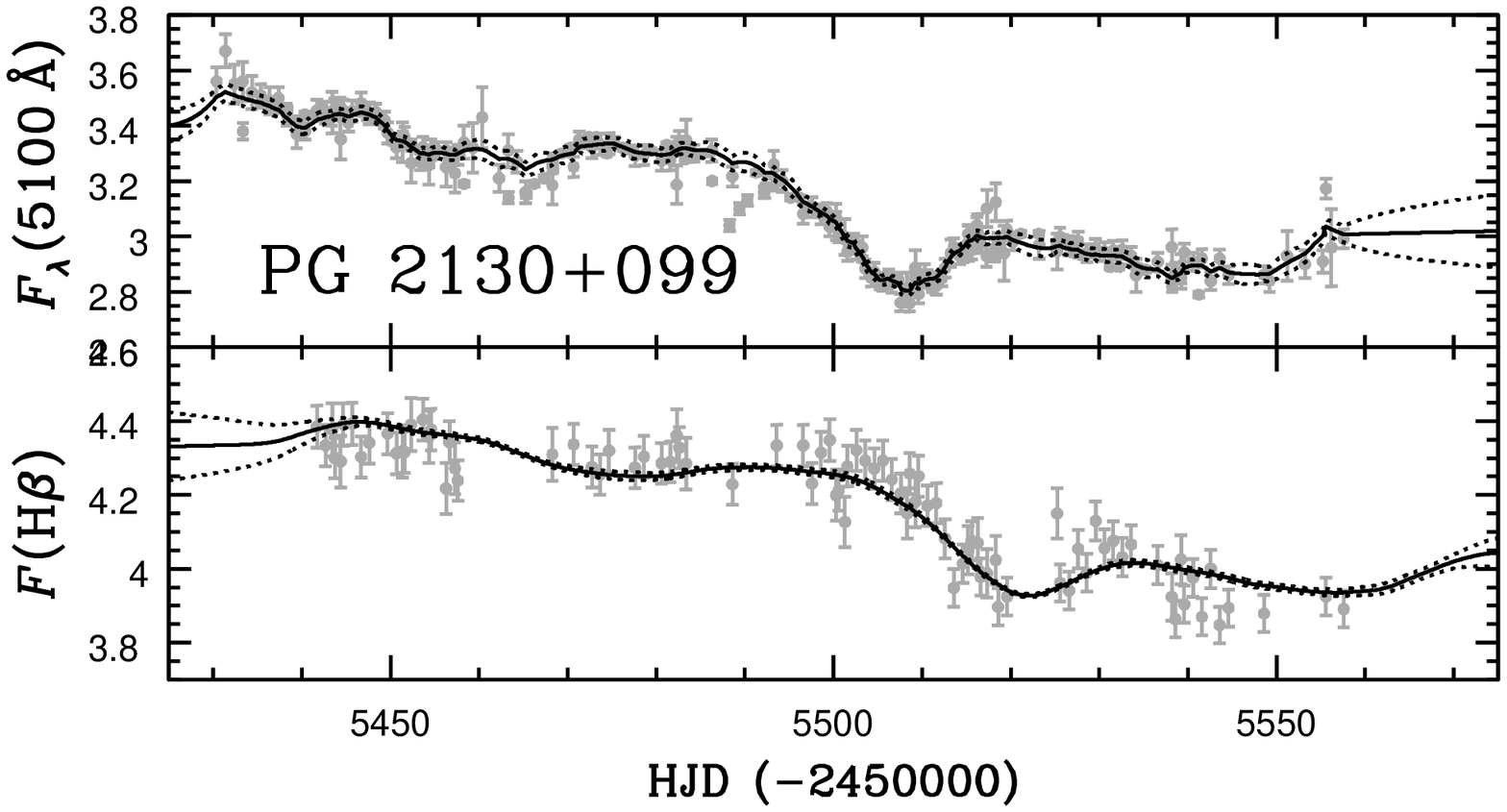}

\caption{{\it Continued.}}
\end{figure}

%%%%%%%%%%%%%%%%%CCFS%%%%%%%%%%%%%%%%%%%%%%
%
%%%%%%%%%%%%%%%%%%Radius-Luminosity Relationship %%%%%%%%%%%%%%%%%%%%%%

\end{document}